\documentclass[a4paper,12pt]{article}

\usepackage{amsmath, amsthm, amsfonts, amssymb, graphicx}

\theoremstyle{plain}
\newtheorem{theorem}{Theorem}[section]

\newtheorem{lemma}[theorem]{Lemma} 

\theoremstyle{remark}

\theoremstyle{definition}

\numberwithin{equation}{section}

\def\tht{\theta}
\def\Om{\Omega}
\def\om{\omega}
\def\e{\varepsilon}
\def\g{\gamma}

\def\l{\lambda}
\def\p{\partial}
\def\D{\Delta}

\def\E{\mbox{\rm e}}
\def\a{\alpha}
\def\b{\beta}

\def\d{\delta}

\def\vp{\varphi}

\def\Odr{\mathcal{O}}
\def\H{W_2}

\def\di{\,\mathrm{d}}

\def\I{\mathrm{I}}
\def\iu{\mathrm{i}}

\DeclareMathOperator{\RE}{Re}

\DeclareMathOperator{\spec}{\sigma}
\DeclareMathOperator{\discspec}{\sigma_{disc}}
\DeclareMathOperator{\essspec}{\sigma_{ess}}

\DeclareMathOperator{\supp}{supp}

\begin{document}

\allowdisplaybreaks

\title{\textbf{On the spectrum of two quantum layers coupled by
a window}}

\author{D.~Borisov}
\date{}
\maketitle

\vspace{-1 true cm}

\begin{quote}
{\small {\em Nuclear Physics Institute, Academy of Sciences,
25068 \v Re\v z
\\
near Prague, Czechia
\\
Bashkir State Pedagogical University, October Revolution St.~3a,
\\
450000 Ufa, Russia
\\
E-mail: \texttt{borisovdi@yandex.ru}}}
\end{quote}

\begin{abstract}
We consider the Dirichlet Laplacian in a domain two
three-dimensional parallel layers having common boundary and
coupled by a window. The window produces the bound states below
the essential spectrum; we obtain two-sided estimates for them.
It is also shown that the eigenvalues emerge from the threshold
of essential spectrum as the window passes through certain
critical shapes. We prove the necessary condition for the window
to be of critical shape. Under an additional assumption we show
that this condition is sufficient and obtain the asymptotic
expansion for the emerging eigenvalue as well as for the
associated eigenfunction.
\end{abstract}

\section*{Introduction}

There is a number of works studying of the spectral properties
the Dirichlet Laplacian in the unbounded domains like infinite
planar strips or three-dimensional layers with some
perturbations. The interest is stimulated by the applications of
such models in quantum mechanics, in particular, in the theory
of quantum waveguides. In the case the perturbation is absent,
the system is trivial due to natural separation of variables,
while the presence usually leads to various phenomena
interesting both from physical and mathematical point of view.

One of the possible system attracting much attention is two
adjacent parallel strips or layers coupled by the window(s)
being bounded domain(s) cut out in the common boundary.  The
two-dimensional case was studied quite intensively, we refer
here to \cite{Bo},  \cite{BE2},  \cite{BEG}, \cite{BGRS},
\cite{ESTV}, \cite{EV2},  \cite{Ga} (see also references
therein). It was shown that the perturbation by the window(s) is
a negative one, i.e., it leads to the presence of the isolated
bound states below the essential spectrum; the latter is
invariant w.r.t. to the window(s). In the case of one window it
was shown in \cite{Bo},  \cite{BEG}, \cite{ESTV} that widening
the window one produces more and more isolated eigenvalues. They
appear when the window's length passes through certain critical
values; this phenomenon was studied in details and the
asymptotics expansions for the emerging eigenvalues were
obtained, see  \cite{Bo}, \cite{BEG}, \cite{Ga}.

In the three-dimensional case corresponding to window-coupled
layers P.~Exner and S.~Vugalter showed that a small window
generates one simple isolated eigenvalue emerging from the
threshold of the essential spectrum \cite{EV2}. They also
obtained two-sided asymptotic estimates for the eigenvalue. The
asymptotics expansion for this eigenvalue has been constructed
formally in \cite{Po}. In the present paper we treat the same
system but for a finite window. The presence of a window leads
to non-empty discrete spectrum; we obtain two-sided estimates
for the eigenvalues. We show that enlargement of the window
produces new isolated eigenvalues emerging from the continuum,
and it happens in the way similar to the two-dimensional case.
Namely, there are critical shapes of the window so that
enlarging the latter one generates a new eigenvalue below the
threshold no matter how the increment is small. We show that the
necessary condition for such eigenvalue to emerge is the
presence of non-trivial bounded resonance solution corresponding
to the threshold of the essential spectrum. 
We describe all possible 
resonance solutions. We also prove that the presence of the
bounded non-trivial resonance solution of certain type is
sufficient to generate an eigenvalue below the essential
spectrum. We also give the leading terms of the asymptotics
expansions for this eigenvalue and the associated eigenfunction.

\section{Formulation of the problem and the main results}

Let $x'=(x_1,x_2)$, $x=(x',x_3)$ be Cartesian coordinates in
$\mathbb{R}^2$ and $\mathbb{R}^3$, respectively, and $\om\subset
\mathbb{R}^2$ be a bounded simply-connected domain having
infinitely differentiable smooth boundary.  We denote
$\Pi_\om:=\{x: x_3\in(-d,0)\cup(0,\pi)\}\cup\om$,
$d\leqslant\pi$. In what follows the set $\om\times\{0\}$ is
referred to as window (cf. Figure). 

The main object of our study is the Dirichlet Laplacian in
$\Pi_\om$ introduced rigorously as associated with the
sesquilinear form
\begin{equation*}
\mathfrak{h}_\om[u,v]:=(\nabla u,\nabla v)_{L_2(\Pi_\om)}
\end{equation*}
on $\overset{0\ }{W_2^1}(\Pi_\om)$, and we indicate it as
$\mathcal{H}_\om$. Hereinafter by $\overset{0\
}{W_2^j}(\Pi_\om)$ we denote the subset of the functions in
$\H^j(\Pi_\om)$ vanishing on $\p\Pi_\om$. Our main aim is to
study the spectrum of $\mathcal{H}_\om$.

In order to present the main results we require additional
notations. Assuming $\om\not=\emptyset$, in a small
neighbourhood of $\p\om$ we introduce coordinates $(\tau,s)$,
where $s$ is the arc length of $\p\om$, and $\tau$ is the
distance to a point measured in the direction of the outward
normal to $\p\om$. By $(r,\tht)$ we denote the polar coordinates
corresponding to $(\tau,x_3)$.

\begin{center}
\includegraphics[width=312 true pt, height=201 true pt]{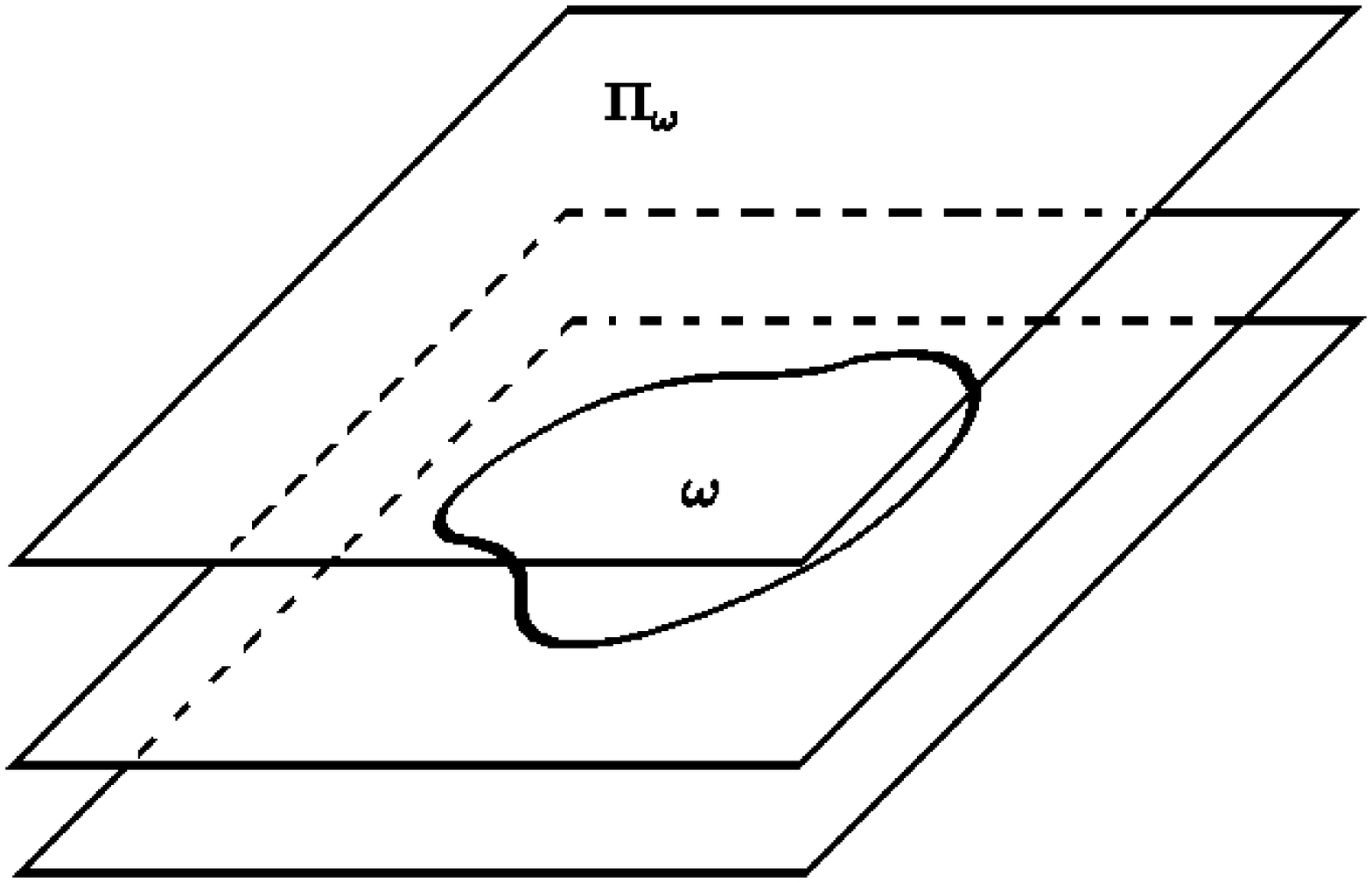}

\medskip

Figure. Window-coupled layers
\end{center}

\medskip

Let $\chi=\chi(t)\in C^\infty(\mathbb{R})$ be a cut-off function
vanishing as $t>1/3$ and equalling one as $t<1/4$. Given $\d>0$
small enough, by $\mathfrak{V}_\d$ we denote the set of the
functions
\begin{gather*}
u(x)=\left(a_0+\sum\limits_{j=1}^{\infty} \E^{-\frac{2\pi j
r}{s_0}} \left(\frac{a_j}{\sqrt{j}}\cos\frac{2\pi j s}{s_0}+
\frac{\widetilde{a}_j}{\sqrt{j}}\sin\frac{2\pi j s}{s_0} \right)
\right)\sqrt{r}\sin\frac{\tht}{2},\quad
\\
a_j, \widetilde{a}_j\in \mathbb{C},\quad
\|u\|_{\mathfrak{V}_\d}^2
:=|a_0|^2+\sum\limits_{j=1}^{\infty} (|a_j|^2+|\widetilde{a}_j|^2)<\infty,
\end{gather*}
defined on $T_\d:=\{x: r<\d\}$. Here $s_0$ is the length of
$\p\om$. We will show that these functions are well-defined (see
Theorem~\ref{th1.1}).

Given $S\subseteq\Pi_{\om}$ and small $\d>0$, by
$\mathfrak{W}(\d,S)$ we denote the set of the functions
\begin{equation}\label{1.1}
u(x)=u^{(0)}(x)\chi\left(\frac{r}{\d}\right)+u^{(1)}(x),
\end{equation}
where $u^{(0)}\in\mathfrak{V}_\d$, $u^{(1)}\in\H^2(S)$, $u=0$ on
$\p\Pi_\om\cap S$. We will employ the symbol
$\mathcal{D}(\cdot)$ to indicate the domain of an operator.

\begin{theorem}\label{th1.1}
Suppose $\om\not=\emptyset$. Then there exists $\d_0>0$ such
that $\mathcal{D}(\mathcal{H}_\om)=\mathfrak{W}(\d,\Pi_\om)$,
and for each $u\in\mathcal{D}(\mathcal{H}_\om)$
\begin{equation}\label{1.1a}
\mathcal{H}_\om u=-2\nabla u^{(0)}\cdot\nabla\chi-
u^{(0)}\D\chi-\D u^{(1)},\quad
\chi=\chi\left(\frac{r}{\d_0}\right)
\end{equation}
The estimates
\begin{gather}
C_1\|\mathcal{H}_\om u\|_{L_2(\Pi_\om)}\leqslant
\|u^{(0)}\|_{\mathfrak{V}_\d}+\|u^{(1)}\|_{W_2^2(\Pi_\om)}\leqslant
C_2\|\mathcal{H}_\om u\|_{L_2(\Pi_\om)}, 
\nonumber
\\
\begin{aligned}
\|u^{(0)}\|_{\H^1(T_\d)}&+\Big\| \frac{\p}{\p s}\nabla
u^{(0)}\Big\|_{L_2(T_\d)} +\Big\|r\frac{\p^2
u^{(0)}}{\p\tau^2}\Big\|_{L_2(T_\d)}
\\
&+\Big\|r\frac{\p^2
u^{(0)}}{\p \tau\p x_3}\Big\|_{L_2(T_\d)} + \Big\|r\frac{\p^2
u^{(0)}}{\p x_3^2}\Big\|_{L_2(T_\d)}\leqslant C
\|u^{(0)}\|_{\mathfrak{V}_\d}.
\end{aligned}\label{1.4}
\end{gather}
hold true, where the constants $C$, $C_i>0$ are independent of
$u^{(0)}$ and $u^{(1)}$.
\end{theorem}

Let $\l_i=\l_i(\om)$ be the isolated eigenvalues of
$\mathcal{H}_\om$ taken counting multiplicity and ordered in the
non-decreasing order. By $\spec(\cdot)$, $\essspec(\cdot)$,
$\discspec(\cdot)$ we denote the spectrum, the essential
spectrum and the discrete one of the operator. We will also use
the symbol $\# A$ to indicate the number of the elements in a
set $A$.

\begin{theorem}\label{th1.2}
The essential spectrum of $\mathcal{H}_\om$ coincides with
$[1,+\infty)$. The discrete spectrum consists of a finite number
of the eigenvalues satisfying inequalities
\begin{equation}\label{1.5}
\frac{\pi^2}{(\pi+d)^2}+\mu_i^{(N)}\leqslant \l_i(\om)\leqslant
\frac{\pi^2}{(\pi+d)^2}+\mu_i^{(D)},
\end{equation}
where $\mu_i^{(N)}$, $\mu_i^{(D)}$ are the eigenvalues of the
Neumann and Dirichlet Laplacian in $\om$, respectively. The
number of the eigenvalues of $\mathcal{H}_\om$ is estimated as
\begin{equation}\label{1.6}
\#\left\{\mu_i^{(D)}: \mu_i^{(D)}<\frac{2\pi d+d^2}{(\pi+d)^2}
\right\}\leqslant\# \discspec(\mathcal{H}_\om) \leqslant
\#\left\{\mu_i^{(N)}: \mu_i^{(N)}<\frac{2\pi d+d^2}{(\pi+d)^2}
\right\}.
\end{equation}
\end{theorem}

We denote $B_\rho(c):=\{x': |x'-c|<\rho\}$.

\begin{theorem}\label{th1.3}

Let $\om=\om(t)\subset \mathbb{R}^2$ be a family of bounded
simply-connected domains having infinitely differentiable
boundary and satisfying the assumption
\begin{enumerate}
\def\theenumi{(A\arabic{enumi})}
\item\label{A2} For each $t_0\in(0,+\infty)$ and $t$ close to
$t_0$ there exist diffeomorphism $\mathcal{M}(t_0,t)\in C^3$
defined in the vicinity of $\om(t_0)$ such that
$\mathcal{M}(t_0,t) \om(t_0)=\om(t)$,
$\mathcal{M}_0(t_0,t_0)=\I$; the components of
$\mathcal{M}(t_0,t)$ and their derivatives up to the second
order are continuous jointly w.r.t. spatial variables and $t$.
\end{enumerate}
Then the eigenvalues of $\mathcal{H}_{\om(t)}$  are continuous
w.r.t. to $t$. If, in addition, the assumption
\begin{enumerate}
\def\theenumi{(A\arabic{enumi})}\setcounter{enumi}{1}

\item\label{A1} There exist $\rho_i=\rho_i(t)$, $i=1,2$, such
that $B_{\rho_1(t)}(0)\subset \om(t)\subset B_{\rho_2(t)}(0)$,
$t\in[0,+\infty)$, and
\begin{equation}\label{1.7}
\lim\limits_{t\to+\infty}\rho_1(t)=+\infty,\quad
\lim\limits_{t\to+0}\rho_2(t)=0;
\end{equation}
$\om(t_1)\subset\om(t_2)$ for all $t_1<t_2$;
\end{enumerate}
holds true, then there exists an infinite sequence
$0=t_1<t_2\leqslant t_3\leqslant \ldots$, such that
$\#\discspec(\mathcal{H}_{\om(t)})=n$, $t\in(t_n,t_{n+1}]$,
$t_n\to+\infty$, $n\to+\infty$, and $\l_n(\om(t))\to 1-0$, $t\to
t_n+0$.
\end{theorem}

This theorem states that there exist critical shapes of $\om$
such that enlarging them one produces new eigenvalue(s) below
the essential spectrum. The next part of our results describes
how such eigenvalues emerge. First we state

\begin{lemma}\label{lm1.4}
The problem
\begin{equation}\label{1.9}
-\D\Psi=\Psi \quad \text{in}\quad\Pi_\om,\qquad \Psi=0\quad
\text{on}\quad\p\Pi_\om,
\end{equation}
has at most finite number of bounded non-trivial solutions
assumed to be even w.r.t. $x_3$ if $d=\pi$. They can be chosen
so that there is at most one solution behaving at infinity as
\begin{equation}\label{1.10}
\Psi=\sin x_3+\Odr(|x'|^{-2}), \quad x'\to+\infty,\quad
x_3\in(0,\pi);
\end{equation}
at most two solutions behaving as
\begin{equation}\label{1.11}
\Psi=\frac{c_1 x_1+c_2 x_2}{|x'|^2}\sin x_3+\Odr(|x'|^{-3}),
\quad x'\to+\infty,\quad x_3\in(0,\pi),\quad |c_1|^2+|c_2|^2=1;
\end{equation}
and a finite number of solutions belonging to $L_2(\Pi_\om)$.
Each of these solutions is infinitely differentiable up to the
boundary except $\p\om\times\{0\}$, while in the vicinity of
$\p\om\times\{0\}$ it behaves as
\begin{equation}\label{1.12}
\Psi(x)=l_{\Psi}(s)\sqrt{r}\sin\frac{\tht}{2}+\Odr(r),\quad
r\to0,
\end{equation}
where $l_{\Psi}\in C^\infty(\p\om)$.
\end{lemma}

Given $\om$, we consider the family of bounded domains
$\om_\e\subset \mathbb{R}^2$ whose boundaries are
$\p\om_\e:=\{x': \tau=\e\b(s)\}$, where $\e\to+0$, and $\b\in
C^\infty(\p\om)$ is an arbitrary function.

\begin{theorem}\label{th1.5}
Suppose the problem (\ref{1.9}) has no bounded non-trivial
solution assumed to be even w.r.t. $x_3$ if $d=\pi$. Then the
operator $\mathcal{H}_{\om_\e}$ has no eigenvalues converging to
one as $\e\to+0$.
\end{theorem}

We introduce two-valued symbol, $\g:=1$, if $d<\pi$, and
$\g:=2$, if $d=\pi$.

\begin{theorem}\label{th1.6}
Suppose the problem (\ref{1.9}) has the unique bounded solution
$\Psi$ assumed to be even w.r.t. $x_3$ if $d=\pi$, and it
satisfies (\ref{1.10}). Then $l_\Psi\not\equiv0$, and there
exists the unique solution $\widetilde{\Psi}$ to (\ref{1.9})
satisfying the conditions
\begin{equation}\label{1.13}
\begin{aligned}
&\widetilde{\Psi}(x)=\frac{l_{\Psi}(s)\b(s)}{2\sqrt{r}}
\sin\frac{\tht}{2}
+l_{\widetilde{\Psi}}(s)\sqrt{r}\sin\frac{\tht}{2}+\Odr(r),\quad
r\to0,
\\
&\widetilde{\Psi}(x)=c\ln |x'|\sin x_3+\Odr(|x'|^{-1}), \quad
|x'|\to+\infty,\quad x_3\in(0,\pi),
\end{aligned}
\end{equation}
where $\widetilde{l}_\Psi\in C^\infty(\p\om)$.  If
\begin{equation}\label{1.14}
\mathfrak{i}_1:=\frac{1}{2\g}\int\limits_{\p\om}\b l_\Psi^2\di
s>0;\quad \text{or}\quad \mathfrak{i}_1=0,\quad
\mathfrak{i}_2:=\frac{1}{2\g}\int\limits_{\p\om}\b l_\Psi
l_{\widetilde{\Psi}}\di s>0,
\end{equation}
then there exists the unique eigenvalue of
$\mathcal{H}_{\om_\e}$ converging to $1-0$ as $\e\to+0$; it is
simple, and
\begin{equation}\label{1.15}
\begin{aligned}
&\l_\e=1-4\E^{-2\mathsf{C}+2\frac{\mathfrak{i}_2}{\mathfrak{i}_1^2}}
\E^{-\frac{2}{\e \mathfrak{i}_1}}\big(1+\Odr(\e)\big),&&
\text{if}\quad \mathfrak{i}_1>0,
\\
&\l_\e=1- \E^{-\frac{2}{\e^2
\mathfrak{i}_2}}\big(c+\Odr(\e)\big), && \text{if}\quad
\mathfrak{i}_1=0,\quad \mathfrak{i}_2>0,
\end{aligned}
\end{equation}
where $c$ is a constant, $\mathsf{C}$ is the Euler constant. The
associated eigenfunction satisfies the identity
\begin{equation}\label{1.16}
\psi_\e=\Psi+\Odr(\sqrt{\e})
\end{equation}
in the norms of $\H^1(S)$ and $\H^2(S\setminus T_\d)$ for each
bounded fixed domain $S\subset\Pi_{\om_\e}$ and $\d>0$. It
decays exponentially at infinity,
\begin{equation*}
\psi_\e=\Odr(\e^{-\sqrt{1-\l_\e}|x'|}|x'|^{-1}),\quad
|x'|\to+\infty.
\end{equation*}
If
\begin{equation}\label{1.18}
\mathfrak{i}_1<0;\quad \text{or}\quad \mathfrak{i}_1=0,\quad
\mathfrak{i}_2<0,
\end{equation}
then the operator $\mathcal{H}_{\om_\e}$ has no eigenvalues
converging to $1-0$ as $\e\to+0$.
\end{theorem}

We observe that the leading terms in the asymptotics
(\ref{1.15}) are discontinuous as $d\to\pi$; this is due to the
presence of $\g$ in the formulas. The similar phenomenon was
found formally in \cite{Po} in the case of small window. We note
that it occurs in two-dimensional case as well, see \cite{Bo}.

Theorem~\ref{th1.5} states that the necessary condition for the
eigenvalues to emerge is the presence of a bounded non-trivial
solution to (\ref{1.9}). There is a number of cases
corresponding to various non-trivial solutions. One of the
possible cases treats theorem~\ref{th1.6}; other cases remain
open. It is an interesting question to obtain the results
similar to Theorem~\ref{th1.6} for the remaining cases. In
particular, we conjecture that the total multiplicity of the
emerging eigenvalues coincides with the number of bounded
non-trivial solutions to (\ref{1.9}). The other conjecture is
that if there exists the unique bounded non-trivial solution to
(\ref{1.9}), then the eigenvalue emerges if $\mathfrak{i}_1>0$,
and does not if $\mathfrak{i}_1<0$. Moreover, if the eigenvalue
emerges, its asymptotics should depend on the behaviour at
infinity of the non-trivial solution. Namely, we conjecture that
\begin{equation}\label{1.19}
\l_\e=1-c\frac{\e}{|\ln\e|}+\ldots,
\end{equation}
if the non-trivial solution satisfies (\ref{1.11}), and
\begin{equation}\label{1.20}
\l_\e=1-c\e+\ldots,
\end{equation}
if the non-trivial solution belongs to $L_2(\Pi)$, where $c$ are
some constants. One of the motivations to these asymptotics is
the results of \cite{KS} where the two-dimensional Schr\"odinger
operator on the plane perturbed by a fast decaying potential was
considered. They addressed the same question on describing the
behaviour of the eigenvalues emerging from the threshold of the
essential spectrum. The asymptotics similar to (\ref{1.15}) were
shown to occur in some cases, while in the other cases the
asymptotics similar to (\ref{1.19}), (\ref{1.20}) were valid.


\section{Domain of $\mathcal{H}_\om$}

The section is devoted to the proof of Theorem~\ref{th1.1}. We
begin with the series of auxiliary lemmas and notations. We
denote $\Om_\d:=\{(\tau,x_3): r<\d\}\setminus\{(\tau,x_3):
x_3=0, \tau>0\}$, where $\d>0$ is small enough.

\begin{lemma}\label{lm2.1}
For each $g\in L_2(\Om_\d)$ there exists the unique generalized
solution $v\in\overset{0\ }{W_2^1}(\Om)$ to
\begin{equation}\label{2.1}
\D_{\tau,x_3}v=g\quad\text{in}\quad \Om,\qquad
v=0\quad\text{on}\quad \p\Om.
\end{equation}
It can be represented as
$v=v^{(0)}+v^{(1)}$, $v^{(0)}=\a\sqrt{r}\sin\frac{\tht}{2}$,
where $v^{(1)}\in\overset{0\ }{W_2^2}(\Om)$. The estimate
\begin{equation}\label{2.3}
|\a|+\|v^{(1)}\|_{\H^2(\Om_\d)}\leqslant C\|g\|_{L_2(\Om_\d)}
\end{equation}
holds true, where the constant $C$ is independent of $g$ and
$\d$.
\end{lemma}

\begin{proof}
It is sufficient to give the proof for two subcases
corresponding to the function $g$ being odd or even w.r.t.
$x_3$. In both cases the unique solvability of (\ref{2.1})
follows from the standard results in theory of generalized
solutions to elliptic boundary value problems.

If $g$ is odd, the generalized solution to (\ref{2.1}) is odd
w.r.t. $x_3$ and hence $v=0$ as $x_3=0$. Thus, this function
solves the boundary value problem like (\ref{2.1}) but in the
half-disk $\Om_\d\cap\{(\tau,x_3): x_3>0\}$. By the smoothness
improving theorems we thus obtain that $v\in\overset{0\
}{W_2^2}(\Om_\d)$, $\a=0$, and the estimate (\ref{2.3}) is
valid.

Suppose now that $g$ is even w.r.t. $x_3$. We expand $g$ into
the Fourier series
\begin{equation*}
g(\tau,x_3)=\sum\limits_{j=0}^{\infty}
g_{2j+1}(r)\sin\frac{2j+1}{2}\tht,\quad g_p(r)=\frac{1}{\pi}
\int\limits_0^{2\pi}g(\tau,x_3)\sin\frac{p\tht}{2}\di\tht,
\end{equation*}
which holds true in $L_2(\Om_\d)$-norm. This fact can be
established by analogy with the proof of Lemma~3.2 in \cite{Bo}.
The Parseval identity
\begin{equation}\label{2.4}
\|g\|_{L_2(\Om)}^2=\pi\sum\limits_{j=0}^{\infty}\int\limits_0^{\d}
|g_{2j+1}(r)|^2r\di r
\end{equation}
is valid. We now solve (\ref{2.1}) by separation of variables,
\begin{gather}
v(\tau,x_3)=\sum\limits_{j=0}^{\infty}
v_{2j+1}(r)\sin\frac{2j+1}{2}\tht,\label{2.5}
\\
v_p(r):=\frac{r^{\frac{p}{2}}}{p}\int\limits_\d^r
t^{-\frac{p}{2}+1}g_p(t)\di t-\frac{r^{-\frac{p}{2}}}{p}
\int\limits_0^r t^{\frac{p}{2}+1}g_p(t)\di t +
\frac{r^{\frac{p}{2}}\d^{-p}}{p} \int\limits_0^\d
t^{\frac{p}{2}+1}g_p(t)\di t. \nonumber
\end{gather}
Let us first analyse the first term in this series. We define
\begin{gather*}
\widetilde{v}_1(r):=v_1(r)-\a\sqrt{r}=r^{\frac{1}{2}}\int\limits_0^r
t^{\frac{1}{2}}g_1(t)\di t-r^{-\frac{1}{2}}\int\limits_0^r
t^{\frac{3}{2}}g_1(t)\di t,
\\
\a:=\frac{1}{\pi}\int\limits_{\Om_\d}
r^{-\frac{1}{2}}\left(\frac{r}{\d}-1\right)g\sin\frac{\tht}{2}\di\tau
\di x_3.
\end{gather*}
It is easy to estimate the constant $\a$:
\begin{equation}\label{2.6}
|\a|^2\leqslant \frac{\|g\|_{L_2(\Om_\d)^2}}{\pi^2}
\int\limits_{\Om_\d} r^{-1}\left(\frac{r}{\d}-1\right)^2\di\tau
\di x_3=\frac{2\d}{3\pi}\|g\|_{L_2(\Om_\d)^2}.
\end{equation}
Employing the estimate
\begin{equation*}
\int\limits_0^r t^{\frac{3}{2}}|g_1(t)|\di t\leqslant
r\int\limits_0^r t^{\frac{1}{2}}|g_1(t)|\di t,
\end{equation*}
we check that
\begin{equation}
\begin{aligned}
\Big\|\widetilde{v}_1(r)\sin\frac{\tht}{2}&\Big\|_{\H^2(\Om_\d)}^2
\leqslant C\int\limits_0^\d \big(|\widetilde{v}_1''|^2
r+|\widetilde{v}_1'|^2 r^{-1}+|\widetilde{v}_1|^2 r^{-3}\big)\di
r
\\
&\leqslant C\int\limits_0^\d r^{-2} \left(\int\limits_0^r
t^{\frac{1}{2}}|g_1(t)|^2\di t\right)^2\di r
\\
&\leqslant C\int\limits_0^\d r^{-\frac{3}{2}}\int\limits_0^r
t^{\frac{3}{2}}|g_1(t)|^2\di t\di r\leqslant C\int\limits_0^\d
r|g_1(r)|^2\di r\leqslant C\|g\|_{L_2(\Om_\d)}^2,
\end{aligned}\label{2.6a}
\end{equation}
where the constant $C$ is independent of $g$ and $\d$. In view
of the inequality obtained and (\ref{2.6}) it remains to show
that the series $\sum\limits_{j=1}^\infty
v_{2j+1}(r)\sin\frac{(2j+1)\tht}{2}$ converges in
$\H^2(\Om_\d)$-norm and to estimate its norm by
$\|g\|_{L_2(\Om_\d)}$. Hence, we should show that
\begin{equation*}
\sum\limits_{j=1}^{\infty}
\left\|v_{2j+1}(r)\sin\frac{(2j+1)\tht}{2}\right\|_{\H^2(\Om_\d)}
\leqslant C\|g\|_{L_2(\Om_\d)}.
\end{equation*}
Employing the definition of $v_p$, (\ref{2.4}), and the estimate
\begin{equation*}
\left|\int\limits_0^\d t^{\frac{p}{2}+1}g_p(t)\di
t\right|^2\leqslant \frac{\d^{p+2}}{p+2}\int\limits_0^\d
r|g_p(r)|^2\di r,
\end{equation*}
we see that it is sufficient to check that
\begin{equation*}
\sum\limits_{j=1}^{\infty} j^2\int\limits_0^\d \left( r^{2j-2}
\left(\int\limits_r^\d t^{-j+\frac{1}{2}} g_j(t)\di t\right)^2+
r^{-2j-4}\left(\int\limits_0^r t^{j+\frac{3}{2}} g_j(t)\di t
\right)^2 \right)\di r\leqslant C\|g\|_{L_2(\Om_\d)}^2,
\end{equation*}
where the constant $C$ is independent of $g$ and $\d$. This
estimate follows from (\ref{2.4}) and the chain of inequalities
\begin{align*}
\int\limits_0^\d r^{2j-2} &\left( \int\limits_r^\d
t^{-j+\frac{1}{2}} g_j(t)\di t\right)^2\di r \leqslant
\int\limits_0^\d\frac{2r^{j+\frac{1}{2}}}{5-2j} \left(1
-\frac{r^{j-\frac{5}{2}}}{\d^{j-\frac{5}{2}}}
\right)\int\limits_r^\d t^{-j-\frac{1}{2}} |g_j(t)|^2\di t\di r
\\
&\leqslant \int\limits_0^\d \frac{2r^{j+\frac{1}{2}}}{|5-2j|}
\int\limits_r^\d t^{-j-\frac{1}{2}} |g_j(t)|^2\di t\di
r\leqslant \frac{4}{|5-2j|(2j+3)}\int\limits_0^\d r|g_j(r)|^2\di
r,
\\
\int\limits_0^\d r^{-2j-4}& \left(\int\limits_0^r
t^{j+\frac{3}{2}} g_j(t)\di t\right)^2 \di r \leqslant
\int\limits_0^\d \frac{2r^{-j-\frac{1}{2}}}{2j+7}\int\limits_0^r
t^{j+\frac{1}{2}}|g_j(t)|^2\di t\\ &\leqslant
\frac{4}{(2j+7)(2j-1)}\int\limits_0^\d r|g_j(r)|^2\di r,
\end{align*}
where we have integrated by parts.
\end{proof}

\begin{lemma}\label{lm2.2}
For each $f\in L_2(T_\d)$ there exists the unique generalized
solution $u\in\overset{0\ }{W_2^1}(T_\d)$  to the problem
\begin{equation}\label{2.7}
\D_{\tau,x_3,s} u=f\quad \text{in}\quad T_\d, \qquad u=0 \quad
\text{on}\quad \p T_\d.
\end{equation}
It can be represented as
\begin{equation}\label{2.7a}
u=u^{(0)}+u^{(1)},\quad u^{(0)}\in\mathfrak{V}_\d,\quad
u^{(1)}\in \H^2(T_\d).
\end{equation}
The estimates (\ref{1.4}) and
\begin{equation}
\|u^{(0)}\|_{\mathfrak{V}_\d} + \|u^{(1)}\|_{\H^2(T_\d)}
\leqslant C\|f\|_{L_2(T_\d)}\label{2.7b}
\end{equation}
are valid.
\end{lemma}

\begin{proof}
The unique solvability of (\ref{2.7}) is obvious. We separate
variables and obtain:
\begin{align}
&f(x)=f_0(\tau,x_3)+\sum\limits_{j=1}^{\infty} \left(
f_j(\tau,x_3)\cos\frac{2\pi j s}{s_0}+
\widetilde{f}_j(\tau,x_3)\sin\frac{2\pi j s}{s_0}
\right),\nonumber
\\
&f_0=\frac{1}{s_0}\int\limits_0^{s_0} f\di s,\quad
f_j=\frac{2}{s_0}\int\limits_0^{s_0} f\cos\frac{2\pi j
s}{s_0}\di s,\quad
\widetilde{f}_j=\frac{2}{s_0}\int\limits_0^{s_0} f\sin\frac{2\pi
j s}{s_0}\di s,\nonumber
\\
&u(x)=u_0(\tau,x_3)+\sum\limits_{j=1}^{\infty} \left(
u_j(\tau,x_3)\cos\frac{2\pi j s}{s_0}+
\widetilde{u}_j(\tau,x_3)\sin\frac{2\pi j s}{s_0}
\right),\label{2.8}
\end{align}
where the series for $f$ converges in $L_2(T_\d)$, and the
coefficients of (\ref{2.8}) are the generalized solutions to
\begin{equation*}
(\D_{\tau,x_3}-N^2) v=g \quad \text{in}\quad \Om_\d,\qquad
v=0\quad \text{on}\quad\p\Om_\d,
\end{equation*}
where $N=2\pi j/s_0$, and $g=f_j$ or $g=\widetilde{f}_j$. These
problems are uniquely solvable in $\overset{0\
}{W_2^1}(\Om_\d)$. By \cite[Ch. V, Sec. 3.5, Eq. (3.16)]{K} and
the identity
\begin{equation*}
\|\nabla
v\|_{L_2(\Om_\d)}^2-N^2\|v\|_{L_2(\Om_\d)}^2=(g,v)_{L_2(\Om_\d)}
\end{equation*}
we have the estimates
\begin{equation}\label{2.10}
\|v\|_{L_2(\Om_\d)}\leqslant
\frac{C}{N^2+1}\|g\|_{L_2(\Om_\d)},\quad
\|v\|_{\H^1(\Om_\d)}\leqslant \frac{C}{N+1}\|g\|_{L_2(\Om_\d)},
\end{equation}
where the constant $C$ is independent of $g$, $N$, and $\d$.
Thus, the series (\ref{2.8}) converges in $\H^1(\Om_\d)$-norm
and therefore gives the generalized solution to (\ref{2.7}).
This solution solves also (\ref{2.1}), where the right-hand side
is $(g+N^2 v)$. We take into account (\ref{2.10}) and apply
Lemma~\ref{lm2.1} to conclude that the function $v$ can be
represented as $v=\a\sqrt{r}\sin\frac{\tht}{2}+v^{(1)}$, where
$\a$ and $v^{(1)}\in\H^2(\Om_\d)$ satisfy (\ref{2.3}).

Let us estimate $\a$ more precisely. It follows from
Lemma~\ref{lm2.1} that the first term in the series (\ref{2.5})
for $v$ satisfies the relations
\begin{equation*}
v_0(r)\sin\frac{\tht}{2}=\a\sqrt{r}\sin\frac{\tht}{2}+
\widetilde{v}_0(\tau,x_3),\quad
v_0(r)=\frac{1}{\pi}\int\limits_0^{2\pi}
v\sin\frac{\tht}{2}\di\tht,
\end{equation*}
where $\widetilde{v}_0\in\H^2(\Om_\d)$. The function $v_0$
solves the problem
\begin{equation*}
\left(\frac{d}{dr}r\frac{d}{dr}-\frac{1}{4r^2}-N^2\right)
v_0=g_0\quad \text{in}\quad (0,\d), \qquad v_0(\d)=0,\qquad
g_0=\frac{1}{\pi}\int\limits_0^{2\pi}
g\sin\frac{\tht}{2}\di\tht.
\end{equation*}
and obeys the condition $v_0(r)\sin\frac{\tht}{2}\in\overset{0\
}{W_2^1}(\Om_\d)$. Hence,
\begin{equation*}
v_0(r)=\int\limits_0^r g_0(t)\frac{\sqrt{t}\sinh
N(r-t)}{N\sqrt{r}}\di t+\frac{\sinh Nr}{N\sqrt{r}\sinh N\d}
\int\limits_0^\d g_0(t)\sqrt{t} \sinh N(t-\d)\di t.
\end{equation*}
Proceeding as in (\ref{2.6a}), we check that
\begin{equation*}
\frac{\sin\frac{\tht}{2}}{N\sqrt{r}}\int\limits_0^r
g_0(t)\sqrt{t} \sinh N(r-t) \di t\in \H^2(\Om_\d),
\end{equation*}
and therefore
\begin{align}
&\a=\int\limits_0^\d \frac{\sqrt{t}\sinh N(t-\d)}{\sinh N\d}
g_0(t)\di t, \nonumber
\\
& |\a|^2\leqslant \int\limits_0^\d \frac{\sinh^2
N(\d-t)}{\sinh^2 N\d}\di t\int\limits_0^\d t |g_0(t)|^2\di t
\leqslant\frac{C}{N}
\|g\|_{L_2(\Om_\d)},\label{2.11}
\end{align}
where $C=\max\limits_{[0,+\infty)} \frac{\sinh 2t-2t}{4\sinh^2
t}<\infty$. It is easy to check that for $N\not=0$
\begin{equation}\label{2.12}
\Big\| \sqrt{r}\E^{-Nr}\sin\frac{\tht}{2}\Big\|_{L_2(\Om_\d)}^2=
\pi\int\limits_0^\d r^2\E^{-2Nr}\di r\leqslant\frac{\pi}{N^3}
\int\limits_0^{+\infty} t^2\E^{-2t}\di t=\frac{\pi}{4N^3}.
\end{equation}
In the same way one can make sure
\begin{equation*}
\Big\|
\sqrt{r}\E^{-Nr}\sin\frac{\tht}{2}\Big\|_{\H^1(\Om_\d)}^2\leqslant
\frac{C}{N},\quad \Big\|
\sqrt{r}(\E^{-Nr}-1)\sin\frac{\tht}{2}\Big\|_{\H^2(\Om_\d)}^2
\leqslant CN,
\end{equation*}
where the constant $C$ is independent of $N$ and $\d$. By
(\ref{2.10}) and (\ref{2.3}) we conclude now that
\begin{align*}
&\Big\|v-\a\sqrt{r}\E^{-Nr}\sin\frac{\tht}{2}\Big\|_{L_2(\Om_\d)}
\leqslant \frac{C}{N^2+1}\|g\|_{L_2(\Om_\d)},
\\
&\Big\|v-\a\sqrt{r}\E^{-Nr}\sin\frac{\tht}{2}\Big\|_{\H^1(\Om_\d)}
\leqslant \frac{C}{N+1}\|g\|_{L_2(\Om_\d)},
\\
&\Big\|v^{(1)}-\a\sqrt{r}(\E^{-Nr}-1)\sin\frac{\tht}{2}
\Big\|_{\H^2(\Om_\d)} \leqslant C\|g\|_{L_2(\Om_\d)},
\end{align*}
where the constant $C$ is independent of $g$, $N$, and $\d$.
These estimates and (\ref{2.11}) applied to the coefficients of
the series (\ref{2.8}) lead us to (\ref{2.7a}), (\ref{2.7b}), if
we denote the fractions $\frac{\a}{\sqrt{N}}$ corresponding to
$u_j$ and $\widetilde{u}_j$ by $a_j$ and $\widetilde{a}_j$. The
inequality (\ref{1.4}) can be checked by estimating the
appropriate norms of $\sqrt{r}\E^{-N r}$ in the same manner as
in (\ref{2.12}).
\end{proof}

\begin{lemma}\label{lm2.3}
There exists ${\d_0}>0$ small enough such that for any $f\in
L_2(T_{\d_0})$ the generalized solution to
\begin{equation}\label{2.17}
\D_x u=f\quad \text{in}\quad T_{\d_0},\qquad
u=0\quad\text{on}\quad\p T_{\d_0},
\end{equation}
satisfies (\ref{2.7a}), (\ref{2.7b}), (\ref{1.4}).
\end{lemma}

\begin{proof}
Lemma~\ref{lm2.2} implies that the domain of the operator
$\D_{\tau,x_3,s}$ in $T_\d$ with Dirichlet boundary conditions
is $\{u: u\in \mathfrak{V}_\d\oplus\H^2(T_\d), u\big|_{\p
T_\d}=0\}$; the action of this operator reads as
$\D_{\tau,x_3,s}^{(D)}u=\D_{\tau,x_3,s}u$.

The Dirichlet Laplacian in $T_\d$ can be written as
\begin{equation*}
\D_x^{(D)}=\D_{\tau,x_3,s}^{(D)}+ \tau\frac{2\mathsf{k}-\tau
\mathsf{k}^2}{(1-\tau \mathsf{k})^2}\frac{\p^2}{\p s^2}+
\mathcal{L}_1 u,\quad \mathcal{L}_1=-\frac{\mathsf{k}}{1-\tau
\mathsf{k}}\frac{\p}{\p\tau}-\frac{\tau \mathsf{k}'}{(1-\tau
\mathsf{k})^3}\frac{\p}{\p s},
\end{equation*}
where $\mathsf{k}=\mathsf{k}(s)\in C^\infty(\p\Om)$. The
operator $\tau\frac{2\mathsf{k}-\tau\mathsf{k}}{(1-\tau
\mathsf{k})^2}\frac{\p^2}{\p s^2}$ is
$\D_{\tau,x_3,s}^{(D)}$-bounded due to (\ref{2.7b}),
(\ref{1.4}), and the bound is estimated by $C\d$, $C$ is
independent of $\d$. The operator $\mathcal{L}_1$ is
$\D_{\tau,x_3,s}^{(D)}$-compact. Employing now \cite[Ch. I\!V,
Sec. 1.1, Th. 1.1]{K}, we conclude that the domain of
$\D_x^{(D)}$ is the same as that of $\D_{\tau,x_3,s}^{(D)}$, if
$\d$ is small enough. Therefore, the representation (\ref{2.7a})
is valid. The estimates (\ref{2.7b}), (\ref{1.4}) for the
solution to (\ref{2.17}) follow from those for the solution to
(\ref{2.7}) and \cite[Ch. I\!V, Sec. 1.4, Th. 1.16]{K}.
\end{proof}

Let $u$ be a function in the domain of $\mathcal{H}_\om$. By the
definition, $u\in\overset{0\ }{W_2^1}(\Pi_\om)$, and it is a
generalized solution to
\begin{equation}\label{2.21}
-\D_x u=f \quad \text{in}\quad \Pi_\om,\qquad u=0\quad
\text{on}\quad \p\Pi_\om,
\end{equation}
where $f=\mathcal{H}_\om u$. Using the smoothness improving
theorems one can make sure that $u\in\H^2(S)$ for any
$S\in\Pi_\om\setminus T_\d$, $\d>0$, and hence
\begin{equation}\label{2.18}
\mathcal{H}_\om u=-\D_{x} u.
\end{equation}
It is also clear that
\begin{equation}\label{2.19}
\|u\|_{\H^1(T_\d)}\leqslant C\|f\|_{L_2(T_\d)}.
\end{equation}
We denote
\begin{equation*}
\widetilde{u}(x):=\left(1-\chi\left(\frac{r}{2\d}\right)\right)
u(x)\in\overset{0\ }{W_2^2}(\Pi_\om).
\end{equation*}
Employing (\ref{2.19}) and proceeding as in the proof of
Lemma~7.1 in \cite[Ch. 3, Sec. 7]{Ld}, one can check that
 and
\begin{equation*}
\|\widetilde{u}\|_{\H^2(\Pi_\om)}\leqslant
C\|f\|_{L_2(\Pi_\om)}.
\end{equation*}
The function
$\widehat{u}(x):=u(x)\chi\left(\frac{r}{2\d}\right)$ is the
solution to (\ref{2.17}) with the right-hand side
\begin{equation*}
\widetilde{f}:=-f-2\nabla_x u\cdot\nabla_x\chi -u\Delta_x
\chi,\quad \chi=\chi\left(\frac{r}{2\d}\right).
\end{equation*}
In view of (\ref{2.19}) we have
$\|\widetilde{f}\|_{L_2(T_\d)}\leqslant C(\d)\|f\|_{L_2(T_\d)}$.
Employing now Lemma~\ref{lm2.3}, we conclude that the
representation (\ref{2.7a}) and the estimates (\ref{2.7b}),
(\ref{1.4}) hold true. It remains to note that by (\ref{2.7a})
\begin{equation*}
\widehat{u}=\chi\left(\frac{r}{\d}\right)\widehat{u}=
\chi\left(\frac{r}{\d}\right)\widehat{u}^{(0)}+
\chi\left(\frac{r}{\d}\right)\widehat{u}^{(1)}.
\end{equation*}
Denoting now
$u^{(1)}:=\widetilde{u}+\chi\left(\frac{r}{\d}\right)\widehat{u}^{(1)}$,
we conclude that (\ref{1.1}) holds true.

If $u$ is given by (\ref{1.1}), it is easy to check that
$u\in\overset{0\ }{W_2^1}(\Pi_\om)$ is the generalized solution
to the problem (\ref{2.21}), where the right-hand side is that
of (\ref{1.1a}). Thus, $u$ belongs to the domain of
$\mathcal{H}_\om$. To prove (\ref{1.1a}), it is sufficient to
substitute (\ref{1.1}) into (\ref{2.18}). The proof of
Theorem~\ref{th1.1} is complete.


\section{Estimates and continuity of the eigenvalues}

In the section we prove Theorems~\ref{th1.2},~\ref{th1.3}.

\begin{proof}[Proof of Theorem~\ref{th1.2}] The main idea of the
proof is borrowed from \cite[Sec. I\!I]{ESTV}. We introduce
additional boundary $\p\om\times(-d,\pi)$  and impose in turn
Dirichlet and Neumann boundary condition on it. As the result,
we have two direct sum $\mathcal{H}_{int}^{(D)}\oplus
\mathcal{H}_{ext}^{(D)}$ and $\mathcal{H}_{int}^{(N)}\oplus
\mathcal{H}_{ext}^{(N)}$, where $\mathcal{H}_{int}^{(D)}$ is the
Dirichlet Laplacian in $\om\times(-d,\pi)$, and
$\mathcal{H}_{ext}^{(D)}$ is the Dirichlet Laplacian in
$\Pi_\om\setminus\big(\om\times(-d,\pi)\big)$. The operators
$\mathcal{H}_{int}^{(N)}$, $\mathcal{H}_{ext}^{(N)}$ are
introduced in the same way; the difference is the boundary
condition on $\p\om\times(-d,\pi)$ which is the Neumann one.

The identities
\begin{equation*}
\essspec(\mathcal{H}_\om)= \spec(\mathcal{H}_{ext}^{(D)})=
\essspec(\mathcal{H}_{ext}^{(D)})=
\spec(\mathcal{H}_{ext}^{(N)})=
\essspec(\mathcal{H}_{ext}^{(N)})=[1,+\infty)
\end{equation*}
can be proven in the same way as the similar identity in the
proof of Theorem~2.1 in \cite{BE2}. The eigenvalues of
$\mathcal{H}_{int}^{(N)}$, $\mathcal{H}_{int}^{(D)}$ are
calculated by separation of variables,
\begin{equation*}
\l_{i,j}(\mathcal{H}_{int}^{(D)})=\mu_i^{(D)}+\frac{\pi^2
j^2}{(\pi+d)^2},\quad
\l_{i,j}(\mathcal{H}_{int}^{(N)})=\mu_i^{(N)}+\frac{\pi^2
j^2}{(\pi+d)^2},\quad i,j\geqslant 1.
\end{equation*}
It is clear that $\pi^2 j^2/(\pi+d)^2>1$, $j\geqslant 2$. Taking
into account this inequality and standard bracketing \cite[Ch.
X\!I\!I\!I, Sec. 15]{RS}, we arrive at the estimates
(\ref{1.5}). The estimates (\ref{1.6}) follow from (\ref{1.5}).
\end{proof}

\begin{proof}[Proof of Theorem~\ref{th1.3}]
Let us prove first that the eigenvalues of
$\mathcal{H}_{\om(t)}$ are continuous w.r.t. $t$. Given
$t_0\in(0,+\infty)$ and $t$ close to $t_0$, we introduce new
variables by the rule
\begin{align*}
&\widetilde{x}=(\widetilde{x}',\widetilde{x}_3),\quad
\widetilde{x}'=\chi\left(\frac{r}{\d}\right)\mathcal{M}(t_0,t)x'+
\left(1-\chi\left(\frac{r}{\d}\right)\right)x',
\\
& \widetilde{x}_3=\left(b(x',t)\chi\left(\frac{r}{\d}\right)+
1-\chi\left(\frac{r}{\d}\right) \right)x_3, \quad
b(x',t)=\sqrt{\left(\frac{\p\widetilde{\tau}}{\p\tau}\right)^2+
\frac{1}{(1-\tau \mathsf{k})^2}\left(\frac{\p\widetilde{s}}{\p
s}\right)^2}
\end{align*}
where $(\widetilde{\tau},\widetilde{s})$ are associated w.r.t.
with $\widetilde{x}'$ in the same way as $(\tau,s)$ and $x'$.
Bearing in mind \ref{A2}, one can easily make sure that the
variables $\widetilde{x}$ are well-defined for $t$ sufficiently
close to $t_0$, and the domain $\Pi_{\om(t_0)}$ is mapped onto
$\Pi_{\om(t)}$ under the change of variables. In the space
$L_2(\Pi_{\om(t_0)})=L_2(\Pi_{\om(t)})$ we define a unitary
operator
\begin{equation*}
(\mathcal{Q}(t)u)=\sqrt{q} u(Q(\cdot)),\quad
q:=\det\left(\frac{\p x_i}{\p
\widetilde{x}_j}\right)_{i,j=1,\ldots,3},
\end{equation*}
where $Q$ is defined as $x=Q(\widetilde{x})$.
 By direct calculations we check that
\begin{align}
&\mathcal{Q}(t)\mathcal{H}_{\om(t)}
\mathcal{Q}^{-1}(t)=b(Q'(\cdot),t)\mathcal{H}_{\om(t_0)}+
\mathcal{L}_2, \label{3.1}
\\
&
\begin{aligned}
\mathcal{L}_2:=&b_{12}\frac{\p^2}{\p\tau\p s}+x_3
b_{13}\frac{\p^2}{\p\tau\p x_3}+b_{22}\frac{\p^2}{\p s^2}+b_{23}
\frac{\p^2}{\p s\p x_3} +x_3 b_{33} \frac{\p^2}{\p x_3^2}
\\
&+b_1\frac{\p}{\p\tau}+b_2\frac{\p}{\p s}+b_3\frac{\p}{\p
x_3}+b_0,
\end{aligned}\label{3.2}
\end{align}
where $b_{i,j}=b_{i,j}(x,t)\in
C\big(T_\d\times(t_0-c,t_0+c)\big)$, $b_i=b_i(x,t)\in
C\big(T_\d\times(t_0-c,t_0+c)\big)$, and
$b_{i,j}(x,0)=b_i(x,0)=0$. The supports of $b_{i,j}$, $b_i$ lie
inside $T_\d$. It is also follows from \ref{A2} that
\begin{equation*}
b(x',t)=1+\widetilde{b}(x',t),\quad \widetilde{b}(x',t)\in
C_0\big(T_\d\times(t_0-c,t_0+c)\big),\quad
\widetilde{b}(x',0)=0.
\end{equation*}
Theorem~\ref{th1.1}, and, in particular, the estimates
(\ref{1.4}) imply that $\mathcal{L}_2$ is
$\mathcal{H}_{\om(t_0)}$-bounded. By (\ref{3.1}) and the last
formula for $b$ we conclude now that the difference
$\big(\mathcal{Q}(t)\mathcal{H}_{\om(t)}\mathcal{Q}^{-1}(t)-
\mathcal{H}_{\om(t_0)}\big)$ is a small perturbation bounded
relatively w.r.t. $\mathcal{H}_{\om(t_0)}$, and
$\big(\mathcal{Q}(t_0)\mathcal{H}_{\om(t_0)}\mathcal{Q}^{-1}(t_0)=
\mathcal{H}_{\om(t_0)}\big)$. Thus, the eigenvalues of
$\mathcal{Q}(t)\mathcal{H}_{\om(t)}\mathcal{Q}^{-1}(t)$, and
hence of $\mathcal{H}_{\om(t)}$ converges to ones of
$\mathcal{H}_{\om(t_0)}$.

Assume now that $\om(t_1)\subset\om(t_2)$ for all $t_1<t_2$.
These are the standard minimax arguments those show that the
eigenvalues $\l_i(\om(t))$ are monotonically decreasing
functions of $t$. Hence, to prove the last statement of the
theorem it is sufficient to show that for each eigenvalue
$\l_i(\om(t))$ there exists $t_i$ such that $\l_i(\om(t))\to
1-0$, as $t\to t_i+0$. Suppose that this is wrong for an
eigenvalue $\l_j(\om(t))$ on a sequence $t^{(m)}\to+0$. In this
case $\l_1(\om(t^{(m)}))\leqslant\l_j(\om(t^{(m)}))\leqslant c
<1$. At the same time, by \cite[Th. 3.1]{EV2}
and the second identity in (\ref{1.7}) we have $\l_1(\om(t))\to
1-0$, $t\to+0$, the contradiction. The sequence of critical
values $t_i$ is infinite due to (\ref{1.5}) and the first
identity in (\ref{1.7}).
\end{proof}


\section{Reduction of the resolvent to a compact operator}

In this section we study the boundary value problem
\begin{equation}\label{4.1}
-\D u=(1-k^2)u+f \quad \text{in}\quad\Pi_\om,\qquad u=0\quad
\text{on}\quad\p\Pi_\om,
\end{equation}
where $k\in\mathbb{C}$ ranges in a small neighbourhood of zero,
$f\in L_2(\Pi_\om)$, $\supp f\subseteq\Pi_{\om,\b}:=\{x:
|x'|<\b\}$, $\b>0$. We choose $\b$ so that $\om\subset\{x':
|x'|<\b/4\}$. If $d=\pi$, we assume in addition that $g$ is even
w.r.t. $x_3$ and the same is for $u$. We should also specify the
behaviour at infinity for the solutions. If $\RE k>0$, we take
the function $u=(\mathcal{H}_\om-1+k^2)^{-1}f$ as the solution
to (\ref{4.1}). For other values of $k$ we will define the
analytic continuation of the operator
$(\mathcal{H}_\om-1+k^2)^{-1}$. We will do it by the technique
employed in \cite[Sec. 3]{Bo}, \cite[Sec. 3.A]{BEG}. We will
also reduce (\ref{4.1}) to a Fredholm equation in
$L_2(\Pi_{\om,\b})$ that will be one of the key ingredient in
the proof of Theorems~\ref{th1.5},~\ref{th1.6}.

Let $g\in L_2(\Pi_{\om,\b})$, $\supp g\subseteq\Pi_{\om,\b}$,
and $g$ is even w.r.t. $x_3$, if $d=\pi$. By $v=v(x,k)$ we
denote the solution to the problem
\begin{equation*}
-\D v=(1-k^2)v+g\quad\text{in}\quad\Pi_\emptyset,\qquad v=0\quad
\text{on}\quad \p\Pi_\emptyset,
\end{equation*}
given by the formulas
\begin{align*}
&v(x,k)=\left\{
\begin{aligned}
& v^+(x,k),\quad x_3\in(0,\pi),
\\
& v^-(x,k),\quad x_3\in(-d,0),
\end{aligned} \right. 
\\
& v^+(x,k)=\sum\limits_{j=1}^{\infty} v_j^+(x',k)\sin j
x_3,\quad v^-(x,k)=\sum\limits_{j=1}^{\infty} v_j^-(x',k)\sin
\frac{\pi j}{d} x_3, 
\\
&v_j^+(x',k):=\frac{1}{2\pi\iu}\int\limits_{\mathbb{R}^2\times(0,\pi)}
g(y)\mathsf{H}_0(\iu|x'-y'|\sqrt{j^2-1+k^2})\sin y_3\di y, 
\\
&v_j^-(x',k):=\frac{1}{2\iu d}
\int\limits_{\mathbb{R}^2\times(-d,0)}
g(y)\mathsf{H}_0\left(\iu|x'-y'|\sqrt{\frac{\pi^2
j^2}{d^2}-1+k^2} \right)\sin \frac{\pi j}{d}y_3\di y, 
\end{align*}
where $\mathsf{H}_0$ is the Hankel function, and $\sqrt{k^2}=k$,
while the other roots are specified by the requirement
$\sqrt{1}=1$. If $k=0$, we introduce the function $v_1^+$ as
\begin{equation*}
v_1^+(x',0)=\frac{1}{\pi^2}\int\limits_{\mathbb{R}^2\times(0,\pi)}
g(y)\ln|x'-y'|\sin y_3\di y,
\end{equation*}
and
\begin{equation*}
v_1^+(x',0)=\frac{1}{\pi^2}\int\limits_{\mathbb{R}^2\times(-\pi,0)}
g(y)\ln|x'-y'|\sin y_3\di y,\quad \text{if}\quad d=\pi.
\end{equation*}
The function $v$ is well-defined and belongs to
$\H^2(\Pi_{\emptyset,\widetilde{\b}})$ for all
$\widetilde{\b}>0$ and considered values of $k$. This fact can
be shown by analogy with the proof of Lemma~3.1 in \cite{Bo}.

Consider the problem
\begin{equation}\label{4.6}
\D w=\D v \quad \text{in}\quad\Pi_\om,\qquad w=v\quad
\text{on}\quad\p\Pi_\om.
\end{equation}
This problem is uniquely solvable in $\H^1(\Pi_{\om,\b})$. We
construct the solution to (\ref{4.1}) as
\begin{equation}\label{4.7}
u(x,k)=\big(\mathcal{A}_1(k)g\big)(x,k):=w(x,k)\chi
\left(\frac{|x'|}{\b}\right)+\left(1-\chi
\left(\frac{|x'|}{\b}\right)\right)v(x,k).
\end{equation}
This function satisfies the boundary condition in (\ref{4.1}).
Substituting it into the equation in (\ref{4.1}), we obtain
\begin{gather}
g+\mathcal{A}_2(k)g=f,\label{4.8}
\\
\mathcal{A}_2(k)g:=(v-w)(\D+1-k^2)\chi
\left(\frac{|x'|}{\b}\right)+2\nabla\chi
\left(\frac{|x'|}{\b}\right)\cdot\nabla(v-w).
\end{gather}

By $\mathfrak{A}$ we denote the set of the operators
$\mathcal{A}=\mathcal{A}(k)$ bounded as ones from
$L_2(\Pi_{\om,\b})$ into $\H^1(\Pi_{\om,\widetilde{\b}})$,
$\mathfrak{W}(\d,\widetilde{\b})$ and
$\H^2(\Pi_{\om,\widetilde{\b}}\setminus T_\d)$ for each
$\widetilde{\b}>0$, $\d>0$, and small $k$, and such that the
function $\mathcal{A}(k)f$ is real-valued for real-valued $f$
and small non-negative $k$.  If an operator $\mathcal{A}(k)$
belongs to $\mathfrak{A}$ and is continuous (uniformly bounded,
holomorphic) w.r.t. $k$ as an operator from $L_2(\Pi_\b)$ into
each of aforementioned spaces, we will say shortly that the
operator $\mathcal{A}(k)$ belongs to $\mathfrak{A}$ and is
continuous (uniformly bounded, holomorphic) w.r.t. $k$.

We denote
\begin{equation*}
\mathsf{a}(g):=\frac{1}{\pi^2} \int\limits_{\Pi_{\om,\b}\cap\{x:
x_3>0 \}} g(x)\sin x_3\di x_3.
\end{equation*}

Repeating the arguments of the proofs of
Lemmas~3.1,~3.3,~3.4,~3.5 in \cite{Bo} and of
Propositions~3.1,~3.2 in \cite{BEG}, and employing
Lemma~\ref{lm2.3} one can prove
\begin{lemma}\label{lm4.1}
Let $k\in\mathbb{C}$ be small enough. The operator
$\mathcal{A}_1(k)\in \mathfrak{A}$ is bounded uniformly w.r.t.
small $k$. The operator $\mathcal{A}_2(k)$ is a linear compact
operator in $L_2(\Pi_{\om,\b})$. For $k\not=0$ it can be
represented as
\begin{equation}\label{4.8a}
\mathcal{A}_1(k)=\mathcal{A}_{2}(k^2)+\mathcal{A}_{3}(k^2)\ln k,
\quad 
\mathcal{A}_2(k)=\mathcal{A}_{5}(k^2)+\mathcal{A}_{6}(k^2)\ln
k,
\end{equation}
where $\mathcal{A}_3(\cdot), \mathcal{A}_4(\cdot)\in
\mathfrak{A}$ are holomorphic, and $\mathcal{A}_5(\cdot)$,
$\mathcal{A}_6(\cdot)$ are linear compact operators in
$L_2(\Pi_{\om,\b})$ being holomorphic w.r.t. $k$. The functions
$\mathcal{A}_i(k^2)f$, $i=5,6$, are real-valued if $f$ is
real-valued and $k^2$ is small and non-negative. For each $f\in
L_2(\Pi_{\om,\b})$ there exists a solution to (\ref{4.1}) given
by $u=\mathcal{A}_1(k)g$. This solution behaves at infinity as
\begin{equation}\label{4.9}
\begin{aligned}
&u(x,k)=\mathsf{c}\left(k,\frac{x'}{|x'|}\right)\E^{-k|x'|}
|x'|^{-\frac{1}{2}} \sin x_3+
\Odr\big(\E^{-k|x'|}|x'|^{-\frac{3}{2}}\big),\quad
x_3\in(0,\pi),
\\
&\mathsf{c}(k,\xi)=-\frac{\sqrt{2\pi}}{4\sqrt{k}}
\mathsf{a}\big( (1+k\xi\cdot
x')g(x)\big)+\Odr(k^{\frac{3}{2}}),\quad k\to0,
\\
&u(x,k)=\Odr\big(
\E^{-\sqrt{\frac{\pi^2}{d^2}-1+k^2}|x'|}|x'|^{-\frac{3}{2}}
\big),\quad x_3\in(-d,0),\quad\text{if}\quad d<\pi,
\end{aligned}
\end{equation}
if $k\not=0$, and
\begin{equation}\label{4.10}
\begin{aligned}
&u(x,0)=\left(\mathsf{a}(g)\ln|x'|+ \frac{c_1 x_1+ c_2 x_2
}{|x'|^2} \right) \sin x_3+\Odr(|x'|^{-2}),\quad x_3\in(0,\pi),
\\
&u(x,0)=\Odr\big(
\E^{-\sqrt{\frac{\pi^2}{d^2}-1}|x'|}|x'|^{-\frac{3}{2}}
\big),\quad x_3\in(-d,0),\quad\text{if}\quad d<\pi.
\end{aligned}
\end{equation}
For each solution to (\ref{4.1}) behaving at infinity in
accordance with (\ref{4.9}), (\ref{4.10})  there exists the
unique solution to (\ref{4.8}) such that $u=\mathcal{A}_1(k)g$.
\end{lemma}

We denote
\begin{equation*}
V_0(x):=\left\{
\begin{aligned}
&\sin x_3, &&  x_3\in(0,\pi),
\\
&(1-\g)\sin x_3,&&  x_3\in(-d,0).
\end{aligned}\right.
\end{equation*}
By $W_0$ we indicate the solution to (\ref{4.6}) as $v=V_0$. We
introduce the functions $U_0$, by (\ref{4.7}) via $V_0$ and
$W_0$. The next lemma is checked by direct calculations.
\begin{lemma}\label{lm4.1a}
The identities
\begin{align*}
\mathcal{A}_{3}(0)g=&(\mathsf{C}-\ln 2)
\mathcal{A}_{4}(0)g+\mathcal{A}_1(0)g,\quad
\mathcal{A}_{4}(0)g= \mathsf{a}(g) U_0, 
\\
\mathcal{A}_{5}(0)g=&(\mathsf{C}-\ln 2)
\mathcal{A}_{6}(0)g+\mathcal{A}_2(0)g,\quad
\mathcal{A}_{6}(0)g=-\mathsf{a}(g)(\D+1)U_0,
%
\end{align*}
hold true.
\end{lemma}

\begin{lemma}\label{lm4.2}
Let $k=0$. There is a finite number of linear independent
non-trivial solutions to (\ref{4.1}), (\ref{4.10}) assumed to be
even w.r.t. to $x_3$ if $d=\pi$. They can be chosen so that
there is at most one solution behaving at infinity
\begin{equation}\label{4.14}
\Psi(x)=\ln|x'|\sin x_3+\Odr(|x'|^{-1}), \quad
|x'|\to+\infty,\quad x_3\in(0,\pi);
\end{equation}
at most two solutions satisfying (\ref{1.11}), and a finite
number of solutions belonging to $L_2(\Pi_\om)$. Each of these
solutions is infinitely differentiable up to the boundary except
$\p\om\times\{0\}$, and in the vicinity of $\p\om\times\{0\}$ it
satisfies (\ref{1.12}).
\end{lemma}

\begin{proof}
The statement on the existence and the number of the solutions
follows immediately from (\ref{4.10}). The claimed smoothness is
due to the standard smoothness improving theorems. The formula
(\ref{1.12}) can be checked by analogy with the proof of
Lemma~4.2 in \cite{Bo}.
\end{proof}

\begin{lemma}\label{lm4.3}
Let $k=0$. The equation (\ref{4.8}) is solvable if and only if
$(f,\Psi_i)_{L_2(\Pi_{\om,\b})}=0$, where $\Psi_i$ are
non-trivial solutions to (\ref{4.1}), (\ref{4.10}). If the
solvability conditions holds true, there exists the unique
solution of (\ref{4.8}) orthogonal to $\phi$.
\end{lemma}

\begin{proof}
By Lemma~\ref{lm4.1} the operator $\mathcal{A}_2(0)$ is compact.
Thus, the equation (\ref{4.8}) is solvable if and only if
$(f,\phi_i^*)_{L_2(\Pi_{\om,\b})}=0$, where $\phi_i^*$ are
non-trivial solutions to the adjoint equation
$\phi_i^*+\mathcal{A}_2^*(0)\phi_i^*=0$. It is sufficient to
show that $\phi_i^*=\Psi_i$. Since the number of $\phi_i^*$ and
$\Psi_i$ are the same, in view of Lemma~\ref{lm4.1} it is
sufficient to check that
\begin{equation*}
0=\big(\Psi_i+\mathcal{A}_2^*(0)\Psi_i,h\big)_{L_2(\Pi_{\om,\b})}=
\big(\Psi_i,h+\mathcal{A}_2(0)h\big)_{L_2(\Pi_{\om,\b})}
\end{equation*}
for all $h\in L_2(\om,\b)$. We denote $u:=\mathcal{A}_1(0)h$; by
the definition of $\mathcal{A}_1(0)$ this function satisfies
(\ref{4.10}). The same formula is valid for $\Psi_i$. Moreover,
$h+\mathcal{A}_2(0)h=-(\D+1)u$. Taking these facts into account
and integrating by parts, we obtain
\begin{align*}
\big(\Psi_i,h+\mathcal{A}_2(0)h\big)_{L_2(\Pi_{\om,\b})}=
-\int\limits_{\Pi_{\om,\b}} \Psi_i(\D+1)u\di
x=-\int\limits_{\Pi_\om}u(\D+1)\Psi_i\di x=0.
\end{align*}
\end{proof}

\begin{proof}[Proof of Lemma~\ref{lm1.4}] The most part of the
lemma follows from Lemma~\ref{lm4.2}; it remains to check the
statement on the solution satisfying (\ref{1.10}). If there
exists the non-trivial solution $u$ behaving at infinity in
accordance with (\ref{4.14}), the problem (\ref{1.9}) can not
has a solution $\Psi$ satisfying (\ref{1.10}). This fact can be
proven by integrating by parts in the integral
$0=\int\limits_{\Om_\b}u(\D+1)\Psi\di x$.

Assume that there is no non-trivial solution obeying
(\ref{4.14}); let us prove that in this case there is the unique
non-trivial solution behaving at infinity as
\begin{equation}\label{4.15}
\Psi(x)=(c\ln|x'|+1)\sin x_3+\Odr(|x'|^{-1}),\quad
|x'|\to+\infty,\quad x_3\in(0,\pi),\quad c\not=0.
\end{equation}
We construct it as
\begin{align*}
&\Psi(x)=\widetilde{\Psi}(x)+\left(1-
\chi\left(\frac{2|x'|}{3\b}\right)\right)\ln|x'|\sin x_3,&&
x_3\in(0,\pi),
\\
&\Psi(x)=\widetilde{\Psi}(x),&& x_3\in(-d,0),\quad\text{if}\quad
d<\pi,
\\
&\Psi(x)=\widetilde{\Psi}(x)+\left(1-
\chi\left(\frac{2|x'|}{3\b}\right)\right)\ln|x'|\sin x_3,&&
x_3\in(-\pi,0),\quad\text{if}\quad d=\pi.
\end{align*}
It leads us to the problem (\ref{4.1}) for $\widetilde{\Psi}$
with $k=0$ where
\begin{align*}
&f(x)=(\D+1)\left(1-
\chi\left(\frac{2|x'|}{3\b}\right)\right)\sin x_3,&&
x_3\in(0,\pi),
\\
&f(x)=(\g-1)(\D+1)\left(1-
\chi\left(\frac{2|x'|}{3\b}\right)\right)\sin x_3,&&
x_3\in(-\pi,0).
\end{align*}
Taking into account Lemma~\ref{lm4.1} and integrating by parts,
it is easy to check that $(f,\Psi_i)_{L_2(\Pi_{\om,\b})}=0$ for
each non-trivial solution $\Psi_i$ of (\ref{4.1}), (\ref{4.10}).
By Lemma~\ref{lm4.3} the equation (\ref{4.8}) is thus solvable
that by Lemma~\ref{lm4.1} proves the solvability of the problem
for $\widetilde{\Psi}$. The uniqueness of $\Psi$ follows from
(\ref{4.15}).
\end{proof}


\section{Singularity of the resolvent}

In this section we study the behaviour of the operator
$(\I+\mathcal{A}_2(k))^{-1}$ in the vicinity of $k=0$. We
remind, that if $d=\pi$, we restrict all the operators on the
even w.r.t. $x_3$ functions. We consider two cases corresponding
to different possibilities of presence of non-trivial solution
described in Lemmas~\ref{lm1.4},~\ref{lm4.2}. The  results of
the section will be employed in the proof of
Theorems~\ref{th1.5},~\ref{th1.6}.

\subsection{Absence of decaying and logarithmically growing non-trivial solution}

Here we deal with the case when the problem (\ref{4.1}),
(\ref{4.10}) has no non-trivial solutions described in
Lemma~\ref{lm4.2}. As it has been shown in the proof of
Lemma~\ref{lm1.4}, in this case  the problem (\ref{1.9}) has the
unique non-trivial solution satisfying (\ref{4.15}).

We substitute (\ref{4.8a}) into (\ref{4.8}) and take into
account Lemma~\ref{lm4.1a}. It leads us to
\begin{equation}\label{5.1}
g+\mathcal{A}_2(0)g-(\ln k-\ln
2+\mathsf{C})\mathsf{a}(g)(\D+1)U_0+ k^2\ln k\mathcal{A}_{7}(k)
g=f,
\end{equation}
where
\begin{equation*}
\mathcal{A}_{7}(k):=\frac{\mathcal{A}_{5}(k^2)-\mathcal{A}_{5}(0)+
\big(\mathcal{A}_{6}(k^2)-\mathcal{A}_{6}(0)\big)\ln k}{k^2\ln
k}
\end{equation*}
is a compact operator in $L_2(\Pi_{\om,\b})$ continuous w.r.t.
small real $k$. The same is true for
$\big(k\mathcal{A}_{7}(k)\big)'$. Hereinafter the expressions
like $\ln k\mathcal{A}_7(k)$ are understood as $(\ln k)\mathcal
{A}_7(k)$.

By the assumption the operator $(\I+\mathcal{A}_2(0))$ is
invertible, and the same is thus true for
$(\I+\mathcal{A}_2(0)+k^2\ln k \mathcal{A}_{7}(k))$. We denote
the inverse to the latter as $\mathcal{A}_{8}(k)$ and apply it
to (\ref{5.1}):
\begin{equation*}
g-(\ln k-\ln 2+\mathsf{C})\mathsf{a}(g) \mathcal{A}_{8}(k)
(\D+1)U_0=\mathcal{A}_{8}(k)f,
\end{equation*}
Now we apply the functional $\mathsf{a}$ to the equation
obtained that solves (\ref{5.1}):
\begin{equation}\label{5.3}
\begin{gathered}
\mathsf{a}(g)=
\frac{\mathsf{a}(\mathcal{A}_{8}(k)f)}{1-(\ln k-
\ln 2+\mathsf{C})
\mathsf{a}\big(\mathcal{A}_{8}(k)(\D+1)U_0\big) },
\\
\big(\I+\mathcal{A}_2(k)\big)^{-1}f
=g
=
\frac{(\ln k- \ln 2+\mathsf{C})\mathsf{a}(\mathcal{A}_{8}(k)f)
\mathcal{A}_{8}(k)(\D+1)U_0}{1-(\ln k-\ln 2+\mathsf{C})
\mathsf{a}\big(\mathcal{A}_{8}(k)(\D+1)U_0\big) }
+\mathcal{A}_{8}(k) f.
\end{gathered}
\end{equation}
Let us prove that the denominator is non-zero. In order to do
it, we need
\begin{lemma}\label{lm5.1}
The identities
\begin{equation*}
\mathcal{A}_1(0)\mathcal{A}_{8}(0)(\D+1) U_0+U_0=\Psi,\quad
\mathsf{a}\big((\D+1)U_0\big)=c,
\end{equation*}
hold true, where $\Psi$ is the unique solution to (\ref{1.9}),
(\ref{1.10}), and $c$ is from (\ref{4.15}).
\end{lemma}

\begin{proof}
It is sufficient to prove the former formula, since it implies
the latter due to (\ref{4.10}).

The definition of $\mathcal{A}_{8}$ yields that
$\mathcal{A}_{8}(0)(\D+1)U_0=(\I+\mathcal{A}_2(0))^{-1}(\D+1)U_0$,
and hence $\mathcal{A}_{8}(0)(\D+1)U_0$ is the function $g$
corresponding to the solution $u$ of (\ref{4.1}), (\ref{4.14})
with $k=0$, $f=(\D+1) U_0$. Thus, the function $u+U_0$ solves
(\ref{1.9}), (\ref{1.10}), that by the uniqueness completes the
proof.
\end{proof}

This lemma and the definition of $\mathcal{A}_{8}$ imply that
\begin{equation*}
\mathsf{a}\big(\mathcal{A}_{8}(k)(\D+1)U_0\big)=c+k^2h(k)\ln k,
\end{equation*}
where the function $h(k)$  is continuous w.r.t. small real $k$,
and the same is true for $\big(k h(k)\big)'$. We substitute this
identity into (\ref{5.3}), take into account (\ref{4.8a}) and
Lemmas~\ref{lm4.1a},~\ref{lm5.1}, and arrive at

\begin{lemma}\label{lm5.4}
If the problem (\ref{1.9}) has no bounded non-trivial solution
satisfying (\ref{4.10}) or (\ref{1.10}), then the operator
$\mathcal{A}_1(k)(\I+\mathcal{A}_2(k))^{-1}\in \mathfrak{A}$ is
bounded uniformly in small real $k$. If the problem (\ref{1.9})
has the unique bounded non-trivial solution and it satisfies
(\ref{1.10}), then
\begin{align*}
\mathcal{A}_1(k)\big(\I+\mathcal{A}_2(k)\big)^{-1}=&(\ln k-\ln
2+\mathsf{C})\mathsf{a}
\big((\I+\mathcal{A}_2(0))^{-1}\cdot\big)\Psi
\\
&+\mathcal{A}_1(0)(\I+\mathcal{A}_2(0))^{-1}+k^2\ln^3 k
\mathcal{A}_{9}(k),
\end{align*}
where $\mathcal{A}_{9}\in \mathfrak{A}$ is continuous w.r.t.
small real $k$ and the same is true for
$\big(k\mathcal{A}_{9}(k)\big)'$.
\end{lemma}

\subsection{Presence of the unique logarithmically growing solution}

In this subsection we study the case when the problem
(\ref{1.9}) has the unique solution satisfying (\ref{4.14}). We
denote it by $\Psi$; let $\phi\in L_2(\Pi_{\om,\b})$ be the
associated solution to (\ref{4.8}) with $k=0$, $f=0$.

We construct the solution to (\ref{4.8}) as
\begin{equation}\label{5.3a}
g=\a\phi+\widetilde{g},
\end{equation}
where $\a=\a(\widetilde{g},f)$ is a constant. We substitute this
identity and (\ref{4.8a}) into (\ref{4.8}) that yields
\begin{equation}\label{5.4}
\begin{aligned}
\widetilde{g}&+\mathcal{A}_2(0)\widetilde{g}-(\ln k-\ln
2+\mathsf{C})\mathsf{a}(\widetilde{g})(\D+1)U_0+ k^2\ln
k\mathcal{A}_{7}(k) \widetilde{g}
\\
&-\a(\widetilde{g},f)(\ln k-\ln
2+C)(\D+1)U_0+\a(\widetilde{g},f) k^2\ln k
\mathcal{A}_{7}(k)\phi=f.
\end{aligned}
\end{equation}
Here we have used the identity
\begin{equation}\label{5.5a}
\mathsf{a}(\phi)=1
\end{equation}
which follows from the definition of $\phi$, and (\ref{4.10}),
(\ref{4.14}). Integrating by parts, one can check that
\begin{equation*}
\int\limits_{\Pi_{\om,\b}}\Psi(\D+1)U_0\di x=-\g \pi^2.
\end{equation*}
Taking into account this formula and Lemma~\ref{lm4.3}, we
calculate the inner product of (\ref{5.4}) and $\Psi$,
\begin{align*}
\g\pi^2(\ln k-\ln 2+\mathsf{C})\big(
\mathsf{a}(\widetilde{g})&+\a(\widetilde{g},f)\big)+k^2
(\mathcal{A}_{7}(k)\widetilde{g},\Psi)_{L_2(\Pi_{\om,\b})}\ln k
\\
&+\a(\widetilde{g},f)
k^2(\mathcal{A}_{7}(k)\phi,\Psi)_{L_2(\Pi_{\om,\b})} \ln
k=(f,\Psi)_{L_2(\Pi_{\om,\b})}.
\end{align*}
Hence,
\begin{equation}
\begin{aligned}
\a(\widetilde{g},f)=&\frac{(f,\Psi)_{L_2(\Pi_{\om,\b})}}{\g\pi^2(\ln
k-\ln 2+\mathsf{C})+k^2\ln k
(\mathcal{A}_{7}(k)\phi,\Psi)_{L_2(\Pi_{\om,\b})} }
\\
&-\frac{\g\pi^2(\ln k-\ln
2+\mathsf{C})\mathsf{a}(\widetilde{g})+k^2
(\mathcal{A}_{7}(k)\widetilde{g},\Psi)_{L_2(\Pi_{\om,\b})}\ln
k}{\g\pi^2(\ln k-\ln 2+\mathsf{C})+k^2
(\mathcal{A}_{7}(k)\phi,\Psi)_{L_2(\Pi_{\om,\b})}\ln k }.
\end{aligned}\label{5.5b}
\end{equation}
We substitute this identity into (\ref{5.4}) and obtain
\begin{gather*}
\widetilde{g}+\mathcal{A}_2(0)\widetilde{g}+k^2\ln k
\mathcal{A}_{10}(k)\widetilde{g}=\widetilde{f},
\\
\begin{aligned}
&\mathcal{A}_{10}(k):=\mathcal{A}_{7}(k)-\frac{\g\pi^2(\ln k-\ln
2+\mathsf{C})
\mathsf{a}(\cdot)+k^2(\mathcal{A}_{7}(k)\cdot,\Psi)_{L_2(\Pi_{\om,\b})}\ln
k}{\g\pi^2(\ln k-\ln 2+\mathsf{C})+k^2
(\mathcal{A}_{7}(k)\phi,\Psi)_{L_2(\Pi_{\om,\b})} \ln
k}\mathcal{A}_{7}(k)\phi
\\
&+\frac{(\mathcal{A}_{7}(k)\cdot,\Psi)_{L_2(\Pi_{\om,\b})}-
\mathsf{a}(\cdot)(\mathcal{A}_{7}(k)\phi,\Psi)_{L_2(\Pi_{\om,\b})}}
{\g\pi^2(\ln k-\ln 2+\mathsf{C})+k^2\ln k
(\mathcal{A}_{7}(k)\phi,\Psi)_{L_2(\Pi_{\om,\b})} }(\ln k-\ln
2+\mathsf{C})(\D+1)U_0,
\end{aligned}
\\
\widetilde{f}=f+ \frac{\big((\ln k-\ln 2+\mathsf{C})(\D+1)
U_0-k^2 \mathcal{A}_{7}(k)\phi\ln k\big)} {\g\pi^2(\ln k-\ln
2+\mathsf{C})+k^2
(\mathcal{A}_{7}(k)\phi,\Psi)_{L_2(\Pi_{\om,\b})}\ln k}
(f,\Psi)_{L_2(\Pi_{\om,\b})}.
\end{gather*}
It is easy to check that the operator $\mathcal{A}_{10}$ is a
compact one in $L_2(\Pi_{\om,\b})$ bounded uniformly in small
real $k$. One can also make sure that
\begin{equation*}
(\widetilde{f},\Psi)_{L_2(\Pi_{\om,\b})}=0,\quad
\big(\mathcal{A}_{10}(k)\widetilde{g},\Psi\big)_{L_2(\Pi_{\om,\b})}=0
\quad \text{for each}\quad \widetilde{g}\in L_2(\Pi_{\om,\b}).
\end{equation*}
Hence, $f\in \{\Psi\}^\bot$, where $\{\Psi\}^\bot$ is the
orthogonal complement to $\Psi$ in $L_2(\Pi_{\om,\b})$. Since
the operator $(\I+\mathcal{A}_2(0))^{-1}: \{\Psi\}^\bot\to
\{\phi\}^\bot$ is bounded by Lemma~\ref{lm4.3}, we conclude that
the operator $\big(\I+\mathcal{A}_2(0)+k^2\ln k
\mathcal{A}_{10}(k)\big)^{-1}\{\Psi\}^\bot\to \{\phi\}^\bot$ is
bounded uniformly in small real $k$. Thus,
\begin{equation*}
\widetilde{g}=\big(\I+\mathcal{A}_2(0)+k^2\ln k
\mathcal{A}_{10}(k)\big)^{-1}. 
\end{equation*}
We also note that by (\ref{5.5a}), (\ref{5.5b})
\begin{align*}
\mathsf{a}(g)=&\a(\widetilde{g})+\mathsf{a}(\widetilde{g},f)=
\frac{(f,\Psi)_{L_2(\Pi_{\om,\b})}}{\g\pi^2(\ln k-\ln
2+\mathsf{C})+k^2\ln k
(\mathcal{A}_{7}(k)\phi,\Psi)_{L_2(\Pi_{\om,\b})} }
\\
&+\frac{k^2 \Big(\mathcal{A}_{7}(k)\big(
\mathsf{a}(\widetilde{g})\phi\ln k
-\widetilde{g}\big),\Psi\Big)_{L_2(\Pi_{\om,\b})}}{\g\pi^2(\ln
k-\ln 2+\mathsf{C})+k^2
(\mathcal{A}_{7}(k)\phi,\Psi)_{L_2(\Pi_{\om,\b})}\ln k }.
\end{align*}
These identities, the formulas for $\a(\widetilde{g},f)$ and
$\widetilde{f}$, (\ref{5.3a}), and Lemma~\ref{lm4.1a} lead us to
\begin{lemma}\label{lm5.3}
Suppose the problem (\ref{1.9}) has the unique non-trivial
solution, and it satisfies (\ref{4.14}). Then the operator
$\mathcal{A}_1(k)\big(\I+\mathcal{A}_2(k)\big)^{-1}\in
\mathfrak{A}$ is bounded uniformly in small real $k$.
\end{lemma}


\section{Eigenvalues emerging from the essential spectrum}

In the section we prove Theorems~\ref{th1.5},~\ref{th1.6}.

\begin{proof}[Proof of Theorem~\ref{th1.5}] Given $\om_\e$, we
describe the domain $\Pi_{\om_\e}$ in terms of the Cartesian
coordinates $\widetilde{x}=(\widetilde{x}',\widetilde{x}_3)$,
and introduce new variables as
\begin{equation}\label{6.0}
x=\chi\left(\frac{\widetilde{r}}{\d}\right) \mathcal{M}(\e)
\widetilde{x}+\left(1-\chi\left(\frac{\widetilde{r}}{\widetilde{\d}}
\right)\right) \widetilde{x},
\end{equation}
where
$\widetilde{r}:=\sqrt{\widehat{\tau}^2+\widetilde{x}_3^2}$,
$(\widetilde{\tau},\widetilde{s})$ are associated with
$\widetilde{x}'$ and $\p\om$, $\widetilde{\d}>0$ is small
enough. The mapping $\mathcal{M}(\e)$ is described by the
formulas
\begin{equation*}
\tau=\widetilde{\tau}-\e\b(s),\quad\widetilde{s}=s,\quad
x_3=\widetilde{x}_3\sqrt{1+\frac{\e^2
(\b'(s))^2}{(1-\widetilde{\tau} \mathsf{k}(s))^2}}.
\end{equation*}
It is clear that under this change of variables the domain
$\Pi_{\om_\e}$ is mapped onto $\Pi_\om$.

We rewrite the eigenvalue equation
$\mathcal{H}_{\om_\e}\psi=\l\psi$ in the variables $x$ that
leads us to
\begin{equation}
\big(\mathcal{H}_{\om}-\e\mathcal{L}_3(\e)\big)\psi=\l\psi,
\label{6.1}
\end{equation}
where $\mathcal{L}_3(\e)$ is given by the expression in the
right-hand side of (\ref{3.2}) with the coefficients belonging
to $C(T_\d\times[0,\e_0])$, $\e_0>0$, and supports lying inside
$T_\d$. The operator $\mathcal{L}_3$ can be represented as
\begin{equation}\label{6.1a}
\begin{aligned}
&\mathcal{L}_3(\e)=\mathcal{L}_4+\e \mathcal{L}_5+\e^2
\mathcal{L}_6(\e), \quad \mathcal{L}_4=\b
\chi\frac{\p}{\p\tau}(\D+1)-(\D+1)\b \chi\frac{\p}{\p\tau},
\\
& \mathcal{L}_5=\frac{1}{2}\b^2\frac{\p}{\p\tau}\chi^2
\frac{\p}{\p\tau}(\D+1)+
\frac{1}{2}(\D+1)\b^2\chi^2\frac{\p^2}{\p\tau^2}-\b\chi
\frac{\p}{\p\tau}(\D+1)\b\chi \frac{\p}{\p\tau}
\\
&\hphantom{L_5=}-\frac{(\b')^2x_3\chi}{2(1-\tau \mathsf{k})^2}
\frac{\p}{\p x_3}(\D+1)+ (\D+1)\frac{(\b')^2x_3\chi}{2(1-\tau
\mathsf{k})^2} \frac{\p}{\p x_3}
\end{aligned}
\end{equation}
where $\b=\b(s)$, $\chi=\chi\left(r/\widetilde{\d}\right)$, and
$\mathcal{L}_6(\e)$ is given by the expression in the right-hand
side of (\ref{3.2}) with the coefficients belonging to
$C(T_\d\times[0,\e_0])$, $\e_0>0$ and supports lying inside
$T_\d$. The operators $\mathcal{L}_4$, $\mathcal{L}_5$ are in
fact second order differential operators satisfying (\ref{3.2})
with some compactly supported continuous coefficients. We write
them in terms of Laplace operators since it is more convenient
for the following arguments. We also observe that the operators
$\mathcal{L}_i$, $i=3,4,5$, are $\mathcal{H}_{\om}$-bounded by
Theorem~\ref{th1.1} and the bounds can be estimated uniformly in
$\e$.

We can rewrite (\ref{6.1}) as
\begin{equation}\label{6.2}
(\mathcal{H}_\om-\l)\psi=\e\mathcal{L}_3\psi.
\end{equation}
If we denote now $\l=1-k^2$, we conclude that an eigenfunction
$\psi$ is a non-trivial solution to (\ref{4.1}), (\ref{4.9})
with $f=f_\e:=\e \mathcal{L}_3\psi$.

Let us find all values of $k$ converging to zero as $\e\to+0$
for which the problem (\ref{4.1}), (\ref{4.9}) with $f=\e
\mathcal{L}_3\psi$ has a non-trivial solution. If the solution
belongs to $\mathcal{D}(\mathcal{H}_\om)$, it will imply that
$\l=1-k^2$ is an eigenvalue of $\mathcal{H}_{\om_\e}$ close to
the threshold of the essential spectrum. In order to find such
values, we employ the approach similar to that used in
\cite{BEG}, \cite{Bo}, \cite{G1}, \cite{G2}.

We note that in the case $d=\pi$ the eigenfunctions of
$\mathcal{H}_{\om_\e}$ are even w.r.t. $x_3$ that can be proved
by analogy with \cite[Lemma~4.1]{Bo}. Because of this in the
case $d=\pi$ we restrict our considerations to even on $x_3$
functions.

It follows from (\ref{6.2}) and Lemma~\ref{lm4.1} that
$\psi=\mathcal{A}_1(k)\big(\I+\mathcal{A}_2(k)\big)^{-1} f_\e$.
We substitute this formula
into (\ref{6.1}) and obtain
\begin{equation}\label{6.4}
f_\e-\e \mathcal{L}_3
\mathcal{A}_1(k)\big(\I+\mathcal{A}_2(k)\big)^{-1} f_\e=0.
\end{equation}
By the hypothesis and Lemmas~\ref{lm1.4},~\ref{lm4.2} in the
case considered the problem (\ref{1.9}), (\ref{4.10}) can have
at most one non-trivial solution, and if exists, it satisfies
(\ref{4.14}). Hence, by Lemmas~\ref{lm5.4},~\ref{lm5.3}, the
estimate (\ref{1.4}), and the definition of $\mathcal{L}_3$ we
conclude that the operator $\mathcal{L}_3
\mathcal{A}_1(k)\big(\I+\mathcal{A}_2(k)\big)^{-1}$ is bounded
uniformly in $\e$ and small real $k$ as an operator in
$L_2(\Pi_{\om,\b})$. Thus, for $\e$ and small real $k$ the
operator $(\I-\e \mathcal{L}_3
\mathcal{A}_1(k)\big(\I+\mathcal{A}_2(k)\big)^{-1})$ is
boundedly invertible, and the equation (\ref{6.4}) has the
trivial solution only. Therefore, the equation (\ref{6.2}) has
no non-trivial solution for small $\e$ and real $k$ that
completes the proof.
\end{proof}

In the proofs of the next theorem we will employ
\begin{lemma}\label{lm6.1}
Suppose that there exists a non-trivial solution $\Psi$ to
(\ref{1.9}), (\ref{4.14}). Then
\begin{equation*}
\mathsf{a}(g)=-\frac{\big((\I+\mathcal{A}_2(0))g,
\Psi\big)_{L_2(\Pi_{\om,\b})}}{\g\pi^2}.
\end{equation*}
\end{lemma}

\begin{proof}
We denote $u:=\mathcal{A}_1(0)g$. This function solves
(\ref{4.1}), (\ref{4.10}) for $k=0$,
$f:=(\I+\mathcal{A}_2(0))g$. Now it is sufficient to integrate
by parts in the integral
$(f,\Psi)_{L_2(\Pi_{\om,\b})}=(f,\Psi)_{L_2(\Pi_\om)}=-\int\limits_\Pi
\Psi(\D+1)u\di x$ to prove the claimed formula.
\end{proof}

\begin{proof}[Proof of Theorem~\ref{th1.6}] We argue here as in
the proof of Theorem~\ref{th1.5} up to the equation (\ref{6.4}).
We substitute the representation for
$\mathcal{A}_1(k)\big(\I+\mathcal{A}_2(k)\big)^{-1}$ given in
Lemma~\ref{lm5.4} into (\ref{6.4}),
\begin{equation}\label{6.5}
\begin{aligned}
f_\e&-\e(\ln k-\ln 2+\mathsf{C})\mathsf{a}
\big((\I+\mathcal{A}_2(0))^{-1}f_\e\big) \mathcal{L}_3\Psi
\\
&-\e \mathcal{L}_3
\mathcal{A}_1(0)(\I+\mathcal{A}_2(0))^{-1}f_\e- \e k^2\ln k
\mathcal{L}_3 \mathcal{A}_9(k)f_\e=0.
\end{aligned}
\end{equation}
By Lemma~\ref{lm5.1} the operator $\mathcal{L}_3\Big(
\mathcal{A}_1(0)(\I+\mathcal{A}_2(0))^{-1}+k^2\ln^3 k
\mathcal{A}_9(k) \Big)$ is bounded uniformly in $\e$ and small
real $k$ as an operator in $L_2(\Pi_\b)$. Hence, the operator
\begin{equation*}
\I-\e \mathcal{L}_3\big(
\mathcal{A}_1(0)(\I+\mathcal{A}_2(0))^{-1}+k^2\ln^3 k
\mathcal{A}_9(k)\big)
\end{equation*}
is boundedly invertible. We denote the inverse by
$\mathcal{A}_{11}(\e,k)$ and apply it to (\ref{6.5}),
\begin{equation}\label{6.6}
f_\e=\e \mathsf{a}\big((\I+\mathcal{A}_2(0))^{-1}f_\e\big)
\mathcal{A}_{11}(\e,k)\mathcal{L}_3\Psi.
\end{equation}
We seek the non-trivial solution to (\ref{6.1}). By
Lemma~\ref{lm4.1} it implies that the associated function $f_\e$
is also non-trivial. Hence, by the identity obtained,
$\mathsf{a}\big((\I+\mathcal{A}_2(0))^{-1}f_\e\big)\not=0$.
Taking this inequality into account, we apply the functional
$\mathsf{a}\big((\I+\mathcal{A}_2(0))^{-1}\cdot\big)$ to
(\ref{6.6}) and arrive at
\begin{equation}\label{6.7}
1=\e(\ln k-\ln 2+\mathsf{C})\mathsf{a}\big(
(\I+\mathcal{A}_2(0))^{-1}\mathcal{A}_{11}(\e,k)\mathcal{L}_3\Psi
\big).
\end{equation}
The roots of this equation are values of $k$ for which the
equation (\ref{6.4}) has a non-trivial solution. This solution
is unique up to a multiplicative constant and reads as follows
\begin{equation}\label{6.8}
f_\e=\e (\ln k-\ln 2+\mathsf{C})
\mathcal{A}_{11}(\e,k)\mathcal{L}_3\Psi.
\end{equation}
The corresponding non-trivial solution to (\ref{4.1}),
(\ref{4.9}) is given by
$\psi_\e=\mathcal{A}_1(k)(\I+\mathcal{A}_2(k))^{-1}f_\e$. In
view of (\ref{6.7}) the coefficient $\mathsf{c}$ in the
asymptotics (\ref{4.9}) satisfies the identity
\begin{align*}
\mathsf{c}&=-\frac{\sqrt{2\pi}}{4\sqrt{k}}\e \mathsf{a}\big(
(\I+\mathcal{A}_2(0))^{-1}\mathcal{A}_{11}(\e,k)\mathcal{L}_3\Psi
\big)+\Odr(1)
\\
&=\frac{\sqrt{2\pi}}{4\sqrt{k}(\ln k-\ln
2+\mathsf{C})}+\Odr(1),\quad k\to+0,
\end{align*}
where we have employed (\ref{6.7}). Hence, $\mathsf{c}$ is
non-zero and the function $\psi_\e$ decays at infinity (and thus
belongs is a needed eigenfunction), if and only if it is
associated with a positive root to (\ref{6.7}). We also note
that for $k=0$ the equation (\ref{6.2}) can not have a
non-trivial solution. Indeed, if so, it satisfies (\ref{4.10}),
that allows us to rewrite (\ref{6.2}) as
\begin{equation*}
f_\e-\e \mathcal{L}_3\mathcal{A}_1(0) (\I+\mathcal{A}_2(0))^{-1}
f_\e=0.
\end{equation*}
By the boundedness of $\mathcal{A}_{11}(\e,0)$ implies $f_\e=0$.

Let us study the existence of positive roots to (\ref{6.7}). We
rewrite the equation (\ref{6.7}) as
\begin{equation}\label{6.9}
\frac{1}{\ln k-\ln 2+\mathsf{C}}-\e\mathsf{a}\big(
(\I+\mathcal{A}_2(0))^{-1}\mathcal{A}_{11}(\e,k)\mathcal{L}_3\Psi
\big)=0.
\end{equation}
The properties of $\mathcal{A}_9$ stated in Lemma~\ref{lm5.4}
and the definition of $\mathcal{A}_{11}$ imply that the function
in the left-hand side of this equation is real-valued and
continuous w.r.t. small non-negative $k$. Moreover,
\begin{equation*}
\left|\frac{d}{dk}\mathsf{a}\big(
(\I+\mathcal{A}_2(0))^{-1}\mathcal{A}_{11}(\e,k)\mathcal{L}_3\Psi
\big) \right|\leqslant C,
\end{equation*}
where the constant $C$ is independent of $\e$ and small
non-negative $k$. Hence, the derivation of the left-hand side in
(\ref{6.9}) is strictly negative for small positive $k$ and
small $\e$. Therefore, this equation has at most one positive
root. It is clear that this root exists, if
\begin{equation}\label{6.10}
\mathsf{a}\big(
(\I+\mathcal{A}_2(0))^{-1}\mathcal{A}_{11}(\e,0)\mathcal{L}_3\Psi
\big)<0,
\end{equation}
and does not exist, if
\begin{equation}\label{6.11}
\mathsf{a}\big(
(\I+\mathcal{A}_2(0))^{-1}\mathcal{A}_{11}(\e,0)\mathcal{L}_3\Psi
\big)>0.
\end{equation}
By Lemma~\ref{lm6.1} and the definition of $\mathcal{A}_{11}$ we
obtain that
\begin{align}
&
\begin{aligned}
&\mathsf{a}\big(
(\I+\mathcal{A}_2(0))^{-1}\mathcal{A}_{11}(\e,0)\mathcal{L}_3\Psi
\big)
\\
&=\mathsf{a}\big( (\I+\mathcal{A}_2(0))^{-1}(\I+\e \mathcal{L}_3
\mathcal{A}_1(0)(\I+\mathcal{A}_2(0))^{-1})\mathcal{L}_3\Psi
\big)+\Odr(\e^2)
\\
&=C_0+\e C_1+\Odr(\e^2),
\end{aligned}\label{6.13}
\\
&C_0:=\frac{(\mathcal{L}_4\Psi,\Psi)_{L_2(\Pi)}}{\g\pi^2},
\quad C_1:=\frac{(\mathcal{L}_5\Psi,\Psi)_{L_2(\Pi_\om)}+
(\mathcal{L}_4\mathcal{A}_1(0)(\I+\mathcal{A}_2(0))^{-1}
\mathcal{L}_4\Psi,\Psi)_{L_2(\Pi_\om)}}{\g\pi^2}.\nonumber
\end{align}
We denote $u:=\mathcal{A}_1(0)(\I+\mathcal{A}_2(0))^{-1}
\mathcal{L}_4\Psi$. It follows from Lemma~\ref{lm4.1} and
(\ref{6.1a}) that $u$ is the unique solution to (\ref{4.1}),
(\ref{4.10}) for $k=0$,
$f=\mathcal{L}_4\Psi=-(\D+1)\b\chi\frac{\p\Psi}{\p\tau}$. We
denote now $\widetilde{\Psi}:=u-\b\chi\frac{\p\Psi}{\p\tau}$ and
conclude that there exists the unique solution to (\ref{1.9})
satisfying (\ref{4.10}) and (\ref{1.13}). The identity
(\ref{1.13}) follows from the formula
\begin{equation*}
\Psi(x)=l_\Psi(s)
r^{1/2}\sin\frac{\tht}{2}+l_{\Psi}^{(1)}(s)r\sin\tht+
l_\Psi^{(2)}(s)r^{3/2}\sin\frac{3\tht}{2}+\Odr(r^2),\quad
r\to+0,
\end{equation*}
$l_\Psi^{(i)}\in C^\infty(\p\om)$, which can be proved by
analogy with Lemma~4.2 in \cite{Bo}. Moreover,
$l_\Psi\not\equiv0$, since otherwise the function
$\frac{\p\Psi}{\p x_1}\in \overset{0\ }{\H^1}(\Pi_\om)$ is a
non-trivial solution to (\ref{1.9}) belonging to $L_2(\Pi_\om)$
that contradicts to the hypothesis. We also note that the
function $u$ satisfies (\ref{1.12}) with $l_\Psi$ replaced by
$l_{\widetilde{\Psi}}$.

We employ now all the aforementioned facts and (\ref{6.1a}), and
integrate by parts,
\begin{align*}
C_0=&-\frac{1}{\g\pi^2}\int\limits_{\Pi_\om}\Psi(\D+1)\b\chi
\frac{\p\Psi}{\p\tau}\di x=-\mathfrak{i}_1,
\\
C_1=&\frac{1}{2\g\pi^2}\int\limits_{\Pi_\om}\Psi
(\D+1)\b^2\chi^2\frac{\p^2\Psi}{\p\tau^2}\di x-
\frac{1}{\g\pi^2}\int\limits_{\Pi_\om}\Psi \b\chi
\frac{\p}{\p\tau}(\D+1)\b\chi \frac{\p\Psi}{\p\tau}\di x
\\
&+\frac{1}{2\g\pi^2}\int\limits_{\Pi_\om}\Psi
(\D+1)\frac{(\b')^2x_3\chi}{2(1-\tau \mathsf{k})^2}
\frac{\p\Psi}{\p x_3}\di
x+\frac{1}{\g\pi^2}\int\limits_{\Pi_\om}\Psi\b\chi\frac{\p}{\p\tau}
(\D+1)u\di x
\\
&-\frac{1}{\g\pi^2}\int\limits_{\Pi_\om}\Psi
(\D+1)\b\chi\frac{\p u}{\p\tau}\di
x=-\frac{1}{\g\pi^2}\int\limits_{\Pi_\om}\Psi
(\D+1)\b\chi\frac{\p u}{\p\tau}\di x=-\mathfrak{i}_2.
\end{align*}
Hence, by (\ref{6.13}),
\begin{equation}\label{6.15}
\mathsf{a}\big(
(\I+\mathcal{A}_2(0))^{-1}\mathcal{A}_{11}(\e,0)\mathcal{L}_3\Psi
\big)=-\mathfrak{i}_1-\e\mathfrak{i}_2+\Odr(\e^2).
\end{equation}
It yields that the inequality (\ref{6.11}) holds true, if the
condition (\ref{1.18}) is valid, i.e., in this case the operator
$\mathcal{H}_{\om_\e}$ has no eigenvalues converging to $1-0$ as
$\e\to+0$. If the condition (\ref{1.14}) is valid, it implies
(\ref{6.10}), and in this case the operator
$\mathcal{H}_{\om_\e}$ has the unique eigenvalue converging to
$1-0$ as $\e\to+0$. This eigenvalues is given by
$\l_\e=1-k_\e^2$, where $k_\e$ is the root to (\ref{6.9}). The
formula (\ref{6.15}) and equation (\ref{6.9}) yield the
asymptotics for $k_\e$,
\begin{align*}
&k_\e=2\E^{-\mathsf{C}+\frac{\mathfrak{i}_2}{\mathfrak{i}_1^2}}
\E^{-\frac{1}{\e \mathfrak{i}_1}}\big(1+\Odr(\e)\big),&&
\text{if}\quad \mathfrak{i}_1>0,
\\
&k_\e=\E^{-\frac{1}{\e^2 \mathfrak{i}_2}}\big(c+\Odr(\e)\big),
&& \text{if}\quad \mathfrak{i}_1=0,\quad \mathfrak{i}_2>0,
\end{align*}
where $c$ is a constant. These formulas prove (\ref{1.15}).

The identities (\ref{6.7}), (\ref{6.8}) and  the representation
for $\mathcal{A}_1(k)(\I+\mathcal{A}_2(k))^{-1}$ given in
Lemma~\ref{lm5.4} imply that

\begin{equation}\label{6.16a}
\psi_\e(x)=\Psi(x)+\Odr(\e)
\end{equation}
in $\H^1(\Pi_{\om,\widetilde{\b}})$ and
$\H^2(\Pi_{\om,\widetilde{\b}}\setminus T_\d)$ for each
$\widetilde{\b}>0$, $\d>0$. Given $\d>0$, we can choose
$\widetilde{\d}$ in (\ref{6.0}) small enough so that
$\widetilde{x}=x$ as
$\widetilde{x}\in\Pi_{\om,\widetilde{\b}}\setminus
T_{\widetilde{\d}}$. Hence, by (\ref{6.16a}) we conclude that
\begin{equation*}
\psi_\e(x(\widetilde{x}))=\Psi(\widetilde{x})+\Odr(\e)
\end{equation*}
in $\H^2(S\setminus T_\d)$ for each fixed bounded domain
$S\subset\Pi_{\om_\e}$ and each $\d>0$.

We pass to the variables $\widetilde{x}$ in (\ref{6.16a}), and
in view of last identity we conclude that the asymptotic
(\ref{1.16}) is valid in the norm $\H^1(\Pi_{\om,\b})$, if
\begin{equation}\label{6.20}
\|\vp_\e \|_{\H^1(T_{\widetilde{\d}})}=\Odr(\sqrt{\e}),\quad
\vp_\e=\vp_\e(\widetilde{x}):=\Psi(x(\widetilde{x}))-\Psi(\widetilde{x}),
\end{equation}
for a fixed $\widetilde{\d}>0$ small enough. Here the norm is
understood in terms of variables $\widetilde{x}$.  The lowest
eigenvalue of Dirichlet Laplacian in $T_{\widetilde{\d}}$
increases unboundedly as $\widetilde{\d}\to+0$. We employ this
fact, the minimax principle, and the obvious identity
$\vp_\e\big|_{\p T_{\widetilde{\d}}}=0$ to conclude that for
$\widetilde{\d}$ small enough the inequality
$\|\nabla\vp_\e\|_{L_2(T_{\widetilde{\d}})}^2
\geqslant 2\|\vp_\e\|_{L_2(T_{\widetilde{\d}})}^2$
holds true. It is also clear that the
$L_2(T_{\widetilde{\d}})$-norm of $\vp_\e$ is bounded uniformly
in $\e$, and
$\|(\D_{\widetilde{x}}+1)\vp_\e\|_{L_2(\widetilde{T}_\d)}=\Odr(\e)$.
We employ two last relations and integrate by parts,
\begin{align*}
&\|\vp_\e\|_{\H^1(T_{\widetilde{\d}})}^2\leqslant 3\left(
\|\nabla_{\widetilde{x}}\vp_\e\|_{L_2(T_{\widetilde{\d}})}^2-
\|\vp_\e\|_{L_2(T_{\widetilde{\d}})}^2 \right)
\\
&= 3\int\limits_{\p T_{\widetilde{\d}}\cap\{x: x_3=0\}} \vp_\e
\left[\frac{\p\vp_\e }{\p \widetilde{x}_3}\right]\di
\widetilde{x}'-3\int\limits_{T_{\widetilde{\d}}}\vp_\e
(\D_{\widetilde{x}}+1)\vp_\e\di \widetilde{x}
\\
&= 3\int\limits_{\p \Pi_{\om}\setminus\p\Pi_{\om_\e}}
\Psi(x(\widetilde{x})) \left[\frac{\p\vp_\e }{\p
\widetilde{x}_3}\right]\di \widetilde{x}'-3\int\limits_{\p
\Pi_{\om_\e}\setminus\p\Pi_{\om}} \Psi(\widetilde{x})
\left[\frac{\p\vp_\e }{\p \widetilde{x}_3}\right]\di
\widetilde{x}'+\Odr(\e),
\end{align*}
where
\begin{equation*}
\left[\frac{\p\vp_\e }{\p
\widetilde{x}_3}\right]:=\frac{\p\vp_\e}{\p
\widetilde{x}_3}(\widetilde{x}',-0)-\frac{\p\vp_\e}{\p
\widetilde{x}_3}(\widetilde{x}',+0).
\end{equation*}
It follows from (\ref{1.12}) that the integrands in the
remaining integrals are bounded uniformly in $\e$. Since the
area of $\p \Pi_{\om}\setminus\p\Pi_{\om_\e}$ and $\p
\Pi_{\om_\e}\setminus\p\Pi_{\om}$ is of order $\e$, we arrive at
the identity (\ref{6.20}).

The exponential decaying of $\psi_\e$ at infinity is due to
(\ref{4.9}).
\end{proof}

\section*{Acknowledgments}

I thank P. Exner, who attracted my attention to the problem
studied in the paper. The part of this work was done during my
visit Universit\'e de Saint-Etienne. I am grateful to the
University and G.~Panasenko for the hospitality extended to me.

The work is supported in parts by RFBR (06-01-00138,
05-01-97912-r\_agidel). The author is also supported by
\emph{Marie Curie International Fellowship} within 6th European
Community Framework (MIF1-CT-2005-006254).





\begin{thebibliography}{99}
\bibitem{Bo} D. Borisov. Discrete spectrum of a pair of
non-symmetric waveguides coupled by a window // Sbornik
Mathematics. 2006. V. 197. No. 4. P. 475-504.

\bibitem{BE2}
D.~Borisov and P.~Exner. Distant perturbation asymptotics in
window-coupled waveguides. I.~The non-threshold case // J. Math.
Phys. 2006. V.  2006. V. 47. No. 11. P. 113502-1 -- 113502-24.

\bibitem{BEG} D.~Borisov, P.~Exner and R.~Gadyl'shin.
Geometric coupling thresholds  in a two-dimensional strip //
Journal of Mathematical Physics. 2002. V. 43. No. 12. P.
6265-6278.


\bibitem{BGRS}
W.~Bulla, F.~Gesztesy, W.~Renger, B.~Simon. Weakly coupled bound
states in quantum waveguides // Proc. Amer. Math. Soc. 1997. V.
125. No. 5. P. 1487-1495.

\bibitem{ESTV} P.~Exner, P.\v Seba, M. Tater, and D. Van\v ek.
Bound states and scattering in quantum waveguides coupled
laterally through a boundary window // J. Math. Phys. 1996. V.
37.  No. 10. P. 4867-4887.


\bibitem{EV2} P.~Exner and S.~Vugalter. Bound-state asymptotic
estimate for window-coupled Dirichlet strips and layers // J.
Phys. A. 1997. V. 30. No. 22. P. 7863-7878.

\bibitem{G1} R. Gadyl'shin.
On local perturbations of Shr\"odinger operator in axis //
Theor. Math. Phys. 2002. V. 132. No. 1. P. 976-982.


\bibitem{G2} R. Gadyl'shin.
On local perturbations of Shr\"odinger operator on the plane
// Theor. Math. Phys. 2004. V. 138. No. 1. P. 33-44.

\bibitem{Ga} R. Gadyl'shin. On regular and singular perturbation
of acoustic and quantum waveguides // Comptes Rendus Mechanique.
2004. V. 332. No. 8. P. 647-652.


\bibitem{Ld} O.A.~Ladyzhenskaya and N.N.~Ural’tseva.
Linear and quasilinear elliptic equations, Academic Press, New
York-London 1968.

\bibitem{K} T.~Kato.
Perturbation theory for linear operators. N.Y.: Springer-Verlag,
1966.

\bibitem{KS} M.~Klaus, and B.~Simon. Coupling constants
threshold in nonrelativistic quantum mechanis. I. Short-range
two-body case // Ann. of Phys. 1980. V. 130. No. 2. P. 251-281.


\bibitem{Po} Yu.~Popov. Asymptotics of bound states and bands
for laterally coupled waveguides and layers // J. Math. Phys.
2002. V. 43. No. 1. P. 215-234.


\bibitem{RS}
M.~Reed, B.~Simon. Methods of Modern Mathematical Physics. I\!V:
Analysis of Operators. Academic Press, New York 1978.



\end{thebibliography}
\end{document}